\documentclass{sig-alternate-05-2015}
\usepackage{graphicx}  
\usepackage{fancyhdr}
\usepackage{lastpage}
\usepackage{tocloft}
\usepackage{lmodern}
\usepackage[utf8]{inputenc}
\usepackage{csvsimple}
\usepackage{amsmath}
\usepackage{mathptmx}
\usepackage{eqnarray}
\usepackage{stmaryrd}
\usepackage{balance}
\usepackage{color}
\usepackage{booktabs}
\usepackage{etoolbox}
\usepackage{subfig}
\usepackage{float}
\usepackage[table,xcdraw]{xcolor}
\usepackage{fixme}
\usepackage{xcolor}
\usepackage{microtype}
\usepackage{pifont}

\fxsetup{
    status=draft,
    author= ,
    layout=inline,
    theme=color}
    
\makeatletter
\def\@copyrightspace{\relax}
\makeatother

\begin{document}

\setcopyright{acmcopyright}

\doi{10.475/123_4}

\isbn{123-4567-24-567/08/06}

\conferenceinfo{PLDI '13}{June 16--19, 2013, Seattle, WA, USA}

\acmPrice{\$15.00}

%
\conferenceinfo{WOODSTOCK}{'97 El Paso, Texas USA}

\title{Breadcrumbs: A Feature Rich Mobility Dataset with\\ Point of Interest Annotation}

%
%
%
%
%

\numberofauthors{1} 
%

\author{
%
%
\alignauthor
Arielle Moro, Vaibhav Kulkarni, Pierre-Adrien Ghiringhelli,\\ Bertil Chapuis,  and Beno\^{i}t Garbinato\\
\vspace{6px}
\email{\{firstname.lastname\}@unil.ch}\\
\affaddr{Distributed Object Programming Laboratory}\\
\affaddr{University of Lausanne, Switzerland}\\
}

\makeatletter

\maketitle

\begin{abstract}

In this paper, we present {\it{Breadcrumbs}}, a mobility dataset collected in the city of Lausanne (Switzerland) from multiple mobile phone sensors (GPS, WiFi, Bluetooth) from~81 users for a duration of~12 weeks.
Currently available mobility datasets are restricted to geospatial information obtained through a single sensor at low spatiotemporal granularities.  
Furthermore, this passively collected data lacks ground-truth information regarding points of interest and their semantic labels. 
These features are critical in order to push the possibilities of geospatial data analysis towards analyzing mobility behaviors and movement patterns at a fine-grained scale.
To this end, {\it{Breadcrumbs}} provides ground-truth and semantic labels for the points of interest of all the participants. 
The dataset also contains fine-grained demographic attributes, contact records, calendar events and social relationship tags between the participants.
In order to demonstrate the significance of the ground-truth annotations, we discuss several use cases of this dataset. 
Furthermore, we compare four contrasting and widely used unsupervised clustering approaches for point of interest extraction from geolocation trajectories.  
Using the ground-truth information, we perform a detailed performance validation of these techniques and highlight their strengths and weaknesses. 
Given that mobility data is derived from an individuals inherent need of participating in activities, narrowing the gap between raw trajectory data points and complete trip annotation in essential. 
We thus make {\it{Breadcrumbs}} accessible to the research community in order to facilitate research in the direction of supervised human mobility learning schemes.

\end{abstract}





\section{Introduction}

\begin{table*}[t!]
\footnotesize
\centering
\resizebox{\textwidth}{!}{%
\begin{tabular}{llllllllllll}
\hline
\textbf{Dataset} & \textbf{\#Participants} & \textbf{Duration} & \textbf{\#Events} & \textbf{sampling rate} & \textbf{Location} & \textbf{GPS} & \textbf{Check-ins} & \textbf{GSM} & \textbf{WiFi} & \textbf{Bluetooth} & \textbf{Annotation} \\ \hline
\rowcolor[HTML]{EFEFEF} 
\textbf{GeoLife}~\cite{geolifepaper} & 178 & 5.5 years & 25M & 5 sec & Beijing & \textcolor{green}{\ding{51}} & \textcolor{red}{\ding{55}} & \textcolor{red}{\ding{55}} & \textcolor{red}{\ding{55}} & \textcolor{red}{\ding{55}} & \textcolor{red}{\ding{55}} \\
\textbf{MDC}~\cite{nmdcpaper} & 185 & 3 years & 11M & - & Lausanne & \textcolor{green}{\ding{51}} & \textcolor{red}{\ding{55}} & \textcolor{green}{\ding{51}} & \textcolor{green}{\ding{51}} & \textcolor{green}{\ding{51}} & relationships \\
\rowcolor[HTML]{EFEFEF} 
\textbf{Privamov}~\cite{privamovpaper} & 100 & 15 months & 15M & - & Lyon & \textcolor{green}{\ding{51}} & \textcolor{red}{\ding{55}} & \textcolor{green}{\ding{51}} & \textcolor{green}{\ding{51}} & \textcolor{red}{\ding{55}} & \textcolor{red}{\ding{55}} \\
\textbf{Reality Mining}~\cite{pentland2009reality} & 100 & 9 months & 5M & - & Boston & \textcolor{red}{\ding{55}} & \textcolor{red}{\ding{55}} & \textcolor{red}{\ding{55}} & \textcolor{red}{\ding{55}} & \textcolor{green}{\ding{51}} & relationships \\
\rowcolor[HTML]{EFEFEF} 
\textbf{FourSquare}~\cite{yang2013fine} & 3112 & 10 months & 9M & - & New York & \textcolor{red}{\ding{55}} & \textcolor{green}{\ding{51}} & \textcolor{red}{\ding{55}} & \textcolor{red}{\ding{55}} & \textcolor{red}{\ding{55}} & relationships \\
\textbf{blebeacon}~\cite{unm-blebeacon-20190312} & 46 & 1 month & 5M & - & California & \textcolor{red}{\ding{55}} & \textcolor{red}{\ding{55}} & \textcolor{red}{\ding{55}} & \textcolor{red}{\ding{55}} & \textcolor{green}{\ding{51}} & \textcolor{red}{\ding{55}} \\
\rowcolor[HTML]{EFEFEF} 
\textbf{hyccups}~\cite{upb-hyccups-20161017} & 72 & 63 days & - & - & Bucharest & \textcolor{red}{\ding{55}} & \textcolor{red}{\ding{55}} & \textcolor{red}{\ding{55}} & \textcolor{green}{\ding{51}} & \textcolor{red}{\ding{55}} & relationships \\
\textbf{sigcomm2009}~\cite{thlab-sigcomm2009-20120715} & 76 & 2 days & - & 120 sec & Barcelona & \textcolor{red}{\ding{55}} & \textcolor{red}{\ding{55}} & \textcolor{red}{\ding{55}} & \textcolor{green}{\ding{51}} & \textcolor{green}{\ding{51}} & \textcolor{red}{\ding{55}} \\
\rowcolor[HTML]{EFEFEF} 
\textbf{telefonica}~\cite{bogomolov2014once} & 342 & 4 weeks & - & - & Spain & \textcolor{red}{\ding{55}} & \textcolor{red}{\ding{55}} & \textcolor{green}{\ding{51}} & \textcolor{red}{\ding{55}} & \textcolor{red}{\ding{55}} & \textcolor{red}{\ding{55}} \\
\textbf{ParticipAct}~\cite{chessa2017mobile} & 300 & 1 year & - & - & Bologna & \textcolor{green}{\ding{51}} & \textcolor{red}{\ding{55}} & \textcolor{red}{\ding{55}} & \textcolor{green}{\ding{51}} & \textcolor{green}{\ding{51}} & \textcolor{red}{\ding{55}} \\
\rowcolor[HTML]{EFEFEF} 
\textbf{Nodobo}~\cite{bell2011nodobo} & 27 & 4 months & 5M & - & Glasgow & \textcolor{red}{\ding{55}} & \textcolor{red}{\ding{55}} & \textcolor{green}{\ding{51}} & \textcolor{green}{\ding{51}} & \textcolor{red}{\ding{55}} & \textcolor{red}{\ding{55}} \\
\textbf{d4d challange}~\cite{furletti2017discovering} & 9M & 1 year & - & - & Senegal & \textcolor{red}{\ding{55}} & \textcolor{red}{\ding{55}} & \textcolor{green}{\ding{51}} & \textcolor{red}{\ding{55}} & \textcolor{red}{\ding{55}} & \textcolor{red}{\ding{55}} \\
\rowcolor[HTML]{EFEFEF} 
\textbf{Gowalla}~\cite{cho2011friendship} & 196,591 & 1.5 years & 6M & - & Worldwide & \textcolor{red}{\ding{55}} & \textcolor{green}{\ding{51}} & \textcolor{red}{\ding{55}} & \textcolor{red}{\ding{55}} & \textcolor{red}{\ding{55}} & relationships \\
\textbf{Brightkite}~\cite{chessa2017mobile} & 58,228 & 1.5 years & 4M & - & Worldwide & \textcolor{red}{\ding{55}} & \textcolor{green}{\ding{51}} & \textcolor{red}{\ding{55}} & \textcolor{red}{\ding{55}} & \textcolor{red}{\ding{55}} & relationships \\
\rowcolor[HTML]{AEB6BF} 
\textbf{Breadcrumbs} & 81 & 12 weeks & 14M & 50 sec & Lausanne & \textcolor{green}{\ding{51}} & \textcolor{red}{\ding{55}} & \textcolor{red}{\ding{55}} & \textcolor{green}{\ding{51}} & \textcolor{green}{\ding{51}} & \begin{tabular}[c]{@{}l@{}}ground truth\\ semantic labels\\ relationships\end{tabular} \\ \hline
\end{tabular}%
}
\caption{A descriptive summary of currently available and widely used geospatial mobility datasets and their features.}
\label{tab:summary_datasets}
\end{table*}

The proliferation of GPS equipped smartphones and Internet connectivity has simplified the process of collecting positional data generated by moving entities. 
Spatiotemporal data streams (trajectories) generated by using people centric sensing technologies~\cite{campbell2008rise} can be stored and made available as mobility datasets.
Modeling human mobility using such datasets is increasingly gaining importance as cities are experiencing rapid transformation and growth, which demands a good understanding of individual mobility behavior.  
Mobility datasets are therefore fundamental for designing and evaluating algorithms pertaining to Geographic Information Systems (GIS) and to facilitate experimental reproducibility. 
More specifically, the advancement of techniques addressing spatiotemporal data-based problems such as predictive queries~\cite{CronenTownsend2002PredictingQP}, object tracking~\cite{Wang2007SimultaneousLM}, trajectory indexing~\cite{Chakka2003IndexingLT}, mobility modeling~\cite{barabasi2005origin} and location privacy~\cite{Shokri2011QuantifyingLP} have transpired due to availability of several geospatial mobility datasets~\cite{geolifepaper, yan2013diversity, pentland2009reality, yang2013fine, privamovpaper, nmdcpaper}.

While the publicly available mobility datasets have been widely used in GIS research to answer classical GIS research questions (movement analysis, trajectory indexing and queries), we highlight four critical limitations that stifle pushing the possibilities of geospatial data analysis further. 
These include: (1) lack of positioning information from multiple sensors, (2) unavailability of geolocation points at a high spatiotemporal granularity throughout the span of data collection duration, (3) lack of ground truth information regarding participant points of interest (POI), and (4) unavailability of semantic information regarding the POI.
Datasets such as~\cite{yang2015modeling, pentland2009reality, yang2013fine} are restricted to traces derived from a single sensor; either GPS, GSM, WiFi or Bluetooth.
Having access to high granularity multi-sensor positioning data can lead complex and richer comparative and compositional studies~\cite{privamovpaper}.
Furthermore, since mobility data is derived from an individual's inherent need of participating in activities, lack of ground truth and semantic information is a critical limitation of available passively collected datasets. 
Currently, large amount of trajectory data is captured, but the associated activity information is poor, which is crucial for research domains such as social network pattern mining~\cite{Eagle2005RealityMS, Cho2011FriendshipAM}, behavioral regularities~\cite{Calabrese2011EstimatingOF}, and behavioral entropy~\cite{kulkarni2019examining}.

In this paper, we present a methodological description of geospatial data collection process to avail a dataset capturing multiple aspects of human mobility behavior. 
We present {\it{Breadcrumbs}}, a mobility dataset containing high granularity geolocation data points from GPS, WiFi, Bluetooth and accelerometer sensors from~81 individuals in Switzerland for a period of 12 weeks. 
We further enrich this dataset with point of interest (POI) ground-truth annotation and semantic labels along with demographic attributes, social relationships, calendar events and contact records.
This information is especially important given that the last decade has seen an increasing demand of understanding the semantic behavior of moving objects in multiple sectors~\cite{Li2010MiningPB}.    
This refers to semantic abstractions of the raw mobility trajectories annotated with the knowledge extracted from the participants. 
Having access to the ground-truth, regarding the places where an individual actually spent time and has a meaningful accord with, is also crucial for corroborating the performance of data mining algorithms.

Along with the dataset description, we frame a research question that highlights the importance of a unique feature of {\it{Breadcrumbs}}, i.e., ground-truth information. 
Our research question is thus: Given the ground-truth information, which clustering approach and parameter settings provides the best sensitivity and specificity results?
We perform a systematic comparison of four widely used spatiotemporal clustering approaches: (1) \emph{k}-means, (1) DBSCAN, (2) DJ Cluster, and (4) DT Cluster. 
We also present our approach to perform accurate validation over the POI ground-truth labels and clusters extracted by different algorithms.
We experimentally demonstrate how the ground-truth information can be used as labels to aid computing the ROC characteristics.

The remainder of the paper is organized as follows.
We present a brief review of the existing mobility datasets and a summary of popular POI extraction techniques in Section~\ref{sec:related_work}.
This Section also presents some use cases facilitated by the {\it{Breadcrumbs}} dataset. 
The data collection process is presented in Section~\ref{sec:data_collection} followed by the quantitative analysis in Section~\ref{sec:quantitative_analysis}.
Then, we present a comparison of four clustering algorithms and precisely describe the way we evaluate them in an evaluation framework in Section~\ref{clustering_comparison}.
We finally conclude the paper in Section~\ref{sec:conclusion}.

\section{Related Work \& Use Cases}
\label{sec:related_work}

In this section, we review the existing mobility datasets and list their features. 
We also present and summarize the spatiotemporal clustering techniques used to extract POIs from the geospatial mobility datasets.
Finally, we highlight some of use cases and fields of research that could benefit from the {\it{Breadcrumbs}} dataset.

\subsection{Mobility Datasets and Applications}

Geolocation mobility datasets mainly contain passively collected positioning data points that form a trajectory~\cite{thlab-sigcomm2009-20120715, furletti2017discovering, bell2011nodobo, geolifepaper, privamovpaper}.
Such datasets focussing explicitly on geo-positioning information have been extensively used in several research domains such as discovery of points of interests~\cite{Vhaduri2016CooperativeDO}, computing trajectory similarity~\cite{Leal2016TowardsAE} and designing frameworks for processing trajectory queries \cite{Zhang2018AGF}.
Another domain of geospatial data research studies the interaction between human mobility behaviors and social relationships.
This was driven by several datasets that provide annotated social relationships with spatiotemporal data points~\cite{nmdcpaper, yang2013fine, cho2011friendship}.  
Such annotated datasets have been used to estimate similarity between users based on their location histories and infer potential social ties. 
The performance of the framework was validated against the ground-truth about social relationship collected from the participants.

The above trends have led to a new domain in location privacy, that focusses on inferring social relationships from mobility datasets.
This task also involves formulating novel inference attacks and improved threat models~\cite{backes2017walk2friends}.
Mobility datasets containing demographic information~\cite{yang2013fine, cho2011friendship, chessa2017mobile} have also contributed to location privacy research, wherein these datasets have been used to construct attacks to infer sensitive demographic attributes~\cite{zhong2015you}.
Furthermore, call detail record datasets with home and work place labels have been used to construct attacks against aggregated mobility datasets~\cite{xu2017trajectory}.

In Table~\ref{tab:summary_datasets}, we present a summary of currently available mobility datasets along with the associated features and the data types. 
We observe that majority of the datasets are restricted to positioning data from either one or two sensors and does not provide demographic attributes or ground-truth information.
MDC dataset offers basic demographic information including the participants sex, age group and gender.
However this information is not available for all the users of the dataset. 
Contrary to the existing datasets, we ensure collection of equivalent data points for each participant, ensuring a satisfactory users to data points ratio.
{\it{Breadcrumbs}} also provides ground-truth annotations and exhaustive demographic attributes as compared to existing mobility datasets as highlighted in Table~\ref{tab:summary_datasets}, necessary to push the boundaries of current research.
This information is critical in order to identify attack vectors and proactively fix the vulnerabilities before releasing user information.
These efforts by the research community have resulted in devising improved strategies for anonymizing and aggregation of user location data~\cite{backes2017walk2friends}. 
Such efforts are also necessary to offer a satisfactory trade-off between utility and privacy of trajectory data~\cite{zhao2019walking}. 
Along these lines, we argue that the annotations provided by the {\it{Breadcrumbs}} dataset with regards to POI ground truth and semantic labelling will foster improvement in constructing such utility/privacy measures.
For instance, formulating a probabilistic model for obfuscation mechanisms, where only the actual points of interests are obfuscated to to maintain a satisfactory trade-off~\cite{elsalamouny2016differential}.

\begin{figure*}[t!]
\centering
\includegraphics[scale=0.38]{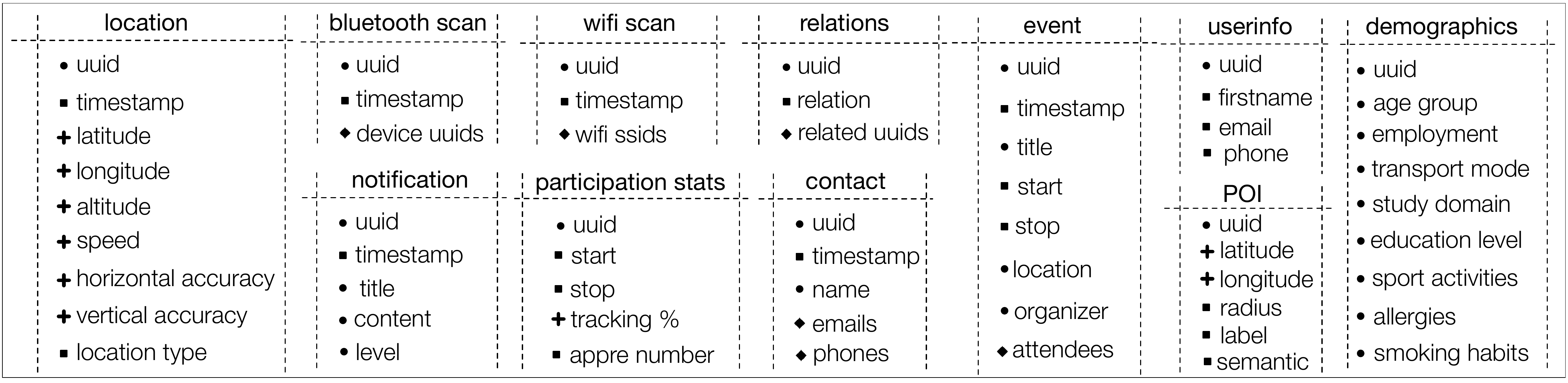}
\caption{Breadcrumbs database schema.}
\label{fig:database_schema}
\vspace{-10pt}
\end{figure*}

\subsection{Point of Interest Extraction}

The seminal work in adopting  data mining and clustering techniques for spatiotemporal POI extracting was proposed by Ashbrook et al.~\cite{Ashbrook2003UsingGT}, where they presented an iterative approach to extract clusters while imposing spatiotemporal bounds.
Then, a two-level clustering approach was proposed by Montoliu et al.~\cite{montoliu2010discovering} wherein the spatiotemporal trajectories are first clustered in the time domain and then in the spatial domain to detect stay-points and successively extract stay-regions.   
Several density based clustering approaches were later proposed such as Density-Joinable clustering~\cite{zhou2004discovering}, Density-Time clustering~\cite{hariharan2004project}, Time-Density Clustering~\cite{gambs2010show} and ZOI detect~\cite{kulkarni:2016:MMP:3003421.3003424}.
These approaches use several parameters to perform clustering in order to extract the POIs.
These parameters include maximum/minimum distance/time between the trajectory data points, cluster shape, maximum number of data points per cluster among others. 
Kulkarni et al.~\cite{Kulkarni:2017:EHW:3139958.3140002} proposed a parameter-less technique for extracting POIs from spatiotemporal trajectories without any {\it{a-priori}} assumptions. 
In this paper, we use a clustering algorithm based on Density-Time clustering (see Section~\ref{ground_truth_information}) to identify hotspots in the dataset participants trajectories. 
We then validated these hotspots through the participants to construct the ground-truth. 
Using this ground-truth information, we compare the performance of clustering technique described in this section. 

\subsection{Research Areas}

In this section, we provide a non-exhaustive list of the research areas and application domains that might benefit from the features provided by {\it{Breadcrumbs}}.\\

\noindent{\bfseries{Next-place Prediction.}} Given the current location of a user, next-place prediction aims at forecasting the place where the user will head next.
POIs used in conjunction with statistical models, such as Markov chains, are known to address this problem reasonably well~\cite{gambs2010show,gambs2012next}.
Other approaches focused on Bayesian networks, neural networks and decision trees as detailed in~\cite{tran2012next,etter2012been}.
However a lot of research recently focused on leveraging recent findings in machine learning to improve next-place predictions.
In this regard, the need to qualitatively compare prediction methods with each other, requires the access to the clusters extracted from the data and the ground truth associated with these clusters.\\

\noindent{\bfseries{Trajectory Prediction.}} Given the current location of a user, trajectory prediction aims at forecasting the trajectory or path that the user will follow while heading to his next POI.
More complex statistical models are required to perform such prediction and the historical GPS traces collected by the user are needed to create more accurate predictions~\cite{chapuis2016capturing}.
In a way which is similar to next-place prediction, the ground-truth is needed to qualitatively assess the forecasts.\\

\noindent{\bfseries{Trajectory Indexing.}} The indexing and retrieval of trajectories and sub-trajectories is a central problem for a wide range of applications, such as car sharing, ride hailing, traffic forecasting, etc.
Recent retrieval techniques focus on the ability to query by trajectories, i.e., the query itself is in the form of a trajectory and the result contain the most relevant matching trajectories of the dataset~\cite{tang2017efficient, chapuis2018geodabs}.
As trajectory indexing can be affected by factors such as the quality of the GPS signal and its sampling rate, a realistic dataset is often necessary to test indexing mechanisms.\\

\noindent{\bfseries{Synthetic Trajectory Generation.}} Location data is often considered as being sensitive.
As a result, sharing them publicly comes with privacy implication and it is difficult to release large and dense datasets made of real trajectories.
Therefore, a common practice consists in creating synthetic but realistic trajectories.
Some models, such as BerlinMod~\cite{duntgen2009berlinmod}, aim at generating trajectories  for benchmarking spatiotemporal databases. 
More recently, machine learning as been used extensively to generate synthetic trajectories.
In this regard, having access to sensor data, such as the accelerometer, can help at generating more realistic trajectories.\\

\noindent{\bfseries{Privacy Preserving LBS.}} As mentioned, sharing location data is often associated with privacy concerns.
For instance, can the demographic group to which a user belong be guessed on the basis of the data recoded by the sensors of his mobile phones?
Having access to a dataset that gathers both location data and demographic data is key to devise and mitigate such inference attacks.
Some research have been designed in this domain in order to infer demographics data by using the entropy level of an individual~\cite{moro2018discovering}.
Other research works also tried to infer demographics information based on other types of data such as access points and location check-ins~\cite{Zhong2015YouAW,Wang2017InferringDA}.
Our dataset is clearly an added value because of the richness of the demographics information.\\

\noindent{\bfseries{Health and Mobility Behavior.}} Several research studies have been done in the domain of health related to mobility behavior.
For example, some researchers studies the mobility of senior individuals~\cite{FILLEKES2019193} and derived indicators based on GPS sensor data.
Other researchers studied the influence of having a dog on the mobility of senior individuals~\cite{thorpe2006dog}.
Finally, \cite{Shoval2008} presented an analysis related to mobility behaviors in the context of cognitive diseases such as Alzheimer.
With the locations of individuals and the additional data captured with the survey, it is possible to conduct health analysis combined with mobility patterns.
Figures~\ref{fig:transportation_modes_weekly_usage_sport_exercise_frequency}, \ref{fig:transportation_modes_weekly_usage_seasonal_allergies}, \ref{fig:transportation_modes_weekly_usage_smokers} and \ref{fig:transportation_modes_weekly_usage_diet} illustrate four examples of research topics related to sport exercise frequency, seasonal allergies, smokers and diet respectively.\\

\noindent{\bfseries{Supervised POI detection.}} User points of interest are currently extracted from mobility trajectories using clustering approaches that are unsupervised in nature. 
We propose a supervised POI detection approach, where the ground truth annotations serve as labels.
In order to formalize the POI classification problem, the raw latitude, longitude and timestamps and passed through a time-series featurizer. 
The featurizer calculates a large number of trajectory characteristics/features such as autocorrelation, entropy, wavelet transform coefficients, number of peaks and crossings among others. 
This set of featurized trajectories can be trained on a binary classifier using the ground truth labels or the semantic labels to predict the next semantic place. 
Such a classifier can be used to asses the feature relevance contributing to a mobility trajectory terminating with a valid POI with a given semantic label.

\newpage
\section{Data Collection Method}
\label{sec:data_collection}

\begin{figure}[h!]
\centering
\includegraphics[scale=0.4]{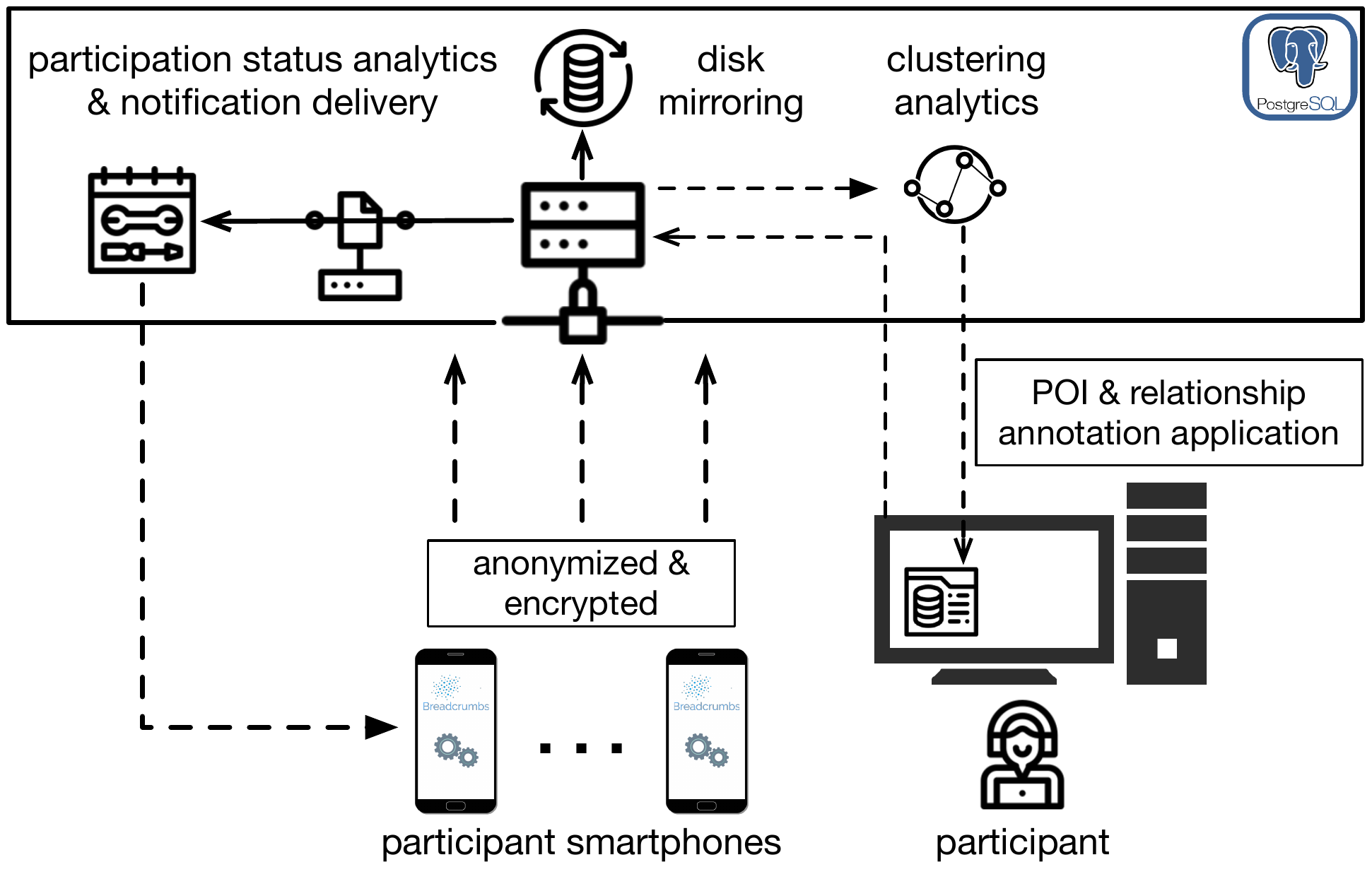}
\caption{Breadcrumbs system architecture.}
\label{fig:breadcrumbs_sys_arch}
\vspace{-10pt}
\end{figure}

The data collection campaign was designed to address the limitations of the currently available spatiotemporal mobility datasets.
In this Section, we describe the design framework adopted to meet each of the limitation and the resulting tradeoffs.
We group the limitations in two categories: (1) collecting high-granularity data points from multiple sensors, and (2) collecting ground-truth information about participant mobility. 
In order to address these limitations, {\it{Breadcrumbs}} data collection campaign aimed at collecting positioning data from GPS, WiFi, Bluetooth and accelerometer sensors, demographic information, calendar events and contact records, ground truth information labelled by the participants at the end of the campaign. 
The dataset contains the above information for~81 individuals, collected mainly in the city of Lausanne (Switzerland) for a period of~12 weeks.
The participants include students from 5 different faculties from universities located in Switzerland and some of their relatives. 
Geolocation information is classified under Personally Identifiable Information (PII) by the EU privacy regulations (GDPR)~\footnote{European Union General Data Protection Regulation:~\url{https://eugdpr.org/}}. 
The system architecture adopted for the {\it{Breadcrumbs}} data collection campaign is presented in Figure~\ref{fig:breadcrumbs_sys_arch}.

\subsection{High Granularity of Multi-Sensor Data}

The data was collected through a mobile application installed on participants' smartphones. 
The application was designed with an objective to be non-intrusive to the participant activity, while optimizing the data collection to power consumption tradeoff.
The sampling rate was calibrated to meet the operational requirements for one day's smartphone usage and obtain a satisfactory granularity in the geolocation data points.
To this end, we limited the sampling of GPS location and accelerometer reading only when the distance between the two location instances is five meters or greater.
The Service Set Identifier (SSID) of the WiFi access point/s and device UUIDs of Bluetooth device/s scan was recorded instantaneously or through the periodic scans by the smartphones   
The data was stored locally, anonymized, encrypted and uploaded to the server when the device is connected to a known WLAN access point.

The server consisted of a PostgreSQL database, a participant notification delivery engine, maintenance and data mirroring frameworks. 
The complete database schema is shown in Figure~\ref{fig:database_schema}. 
The periodic analysis based on this data was used to send the participation status notifications to the respective individual.
This system was also used to recommend data collection and power usage improvement strategies to ensure an uninterrupted data stream.

\subsection{Ground-Truth Information}\label{ground_truth_information}

\begin{figure}[h!]
\centering
\begin{minipage}{.48\linewidth}
  \includegraphics[width=\linewidth]{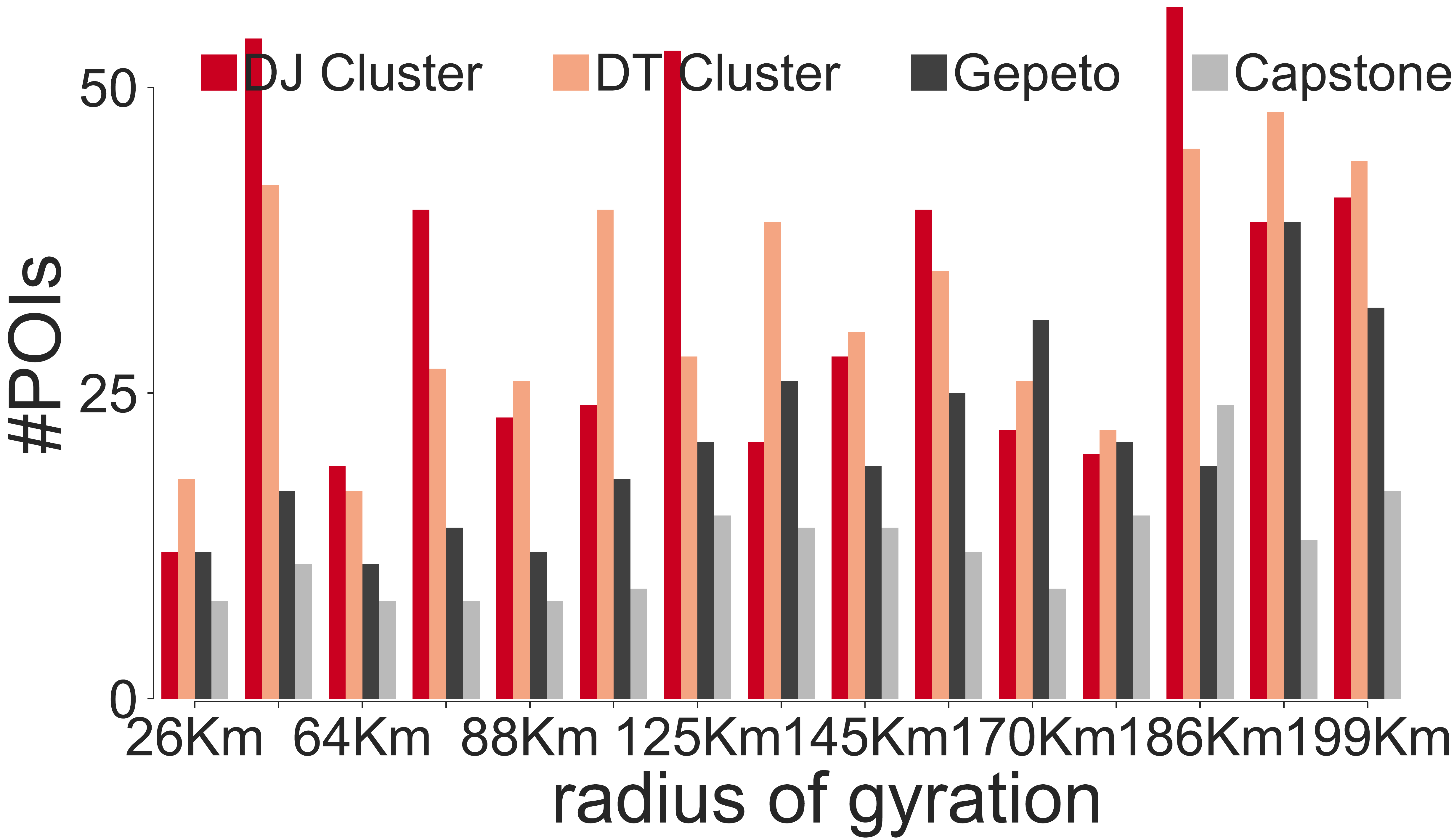}
  \captionof{figure}{MDC Dataset}
  \label{fig:mdc_benchmark}
\end{minipage}
\hspace{.01\linewidth}
\begin{minipage}{.48\linewidth}
  \includegraphics[width=\linewidth]{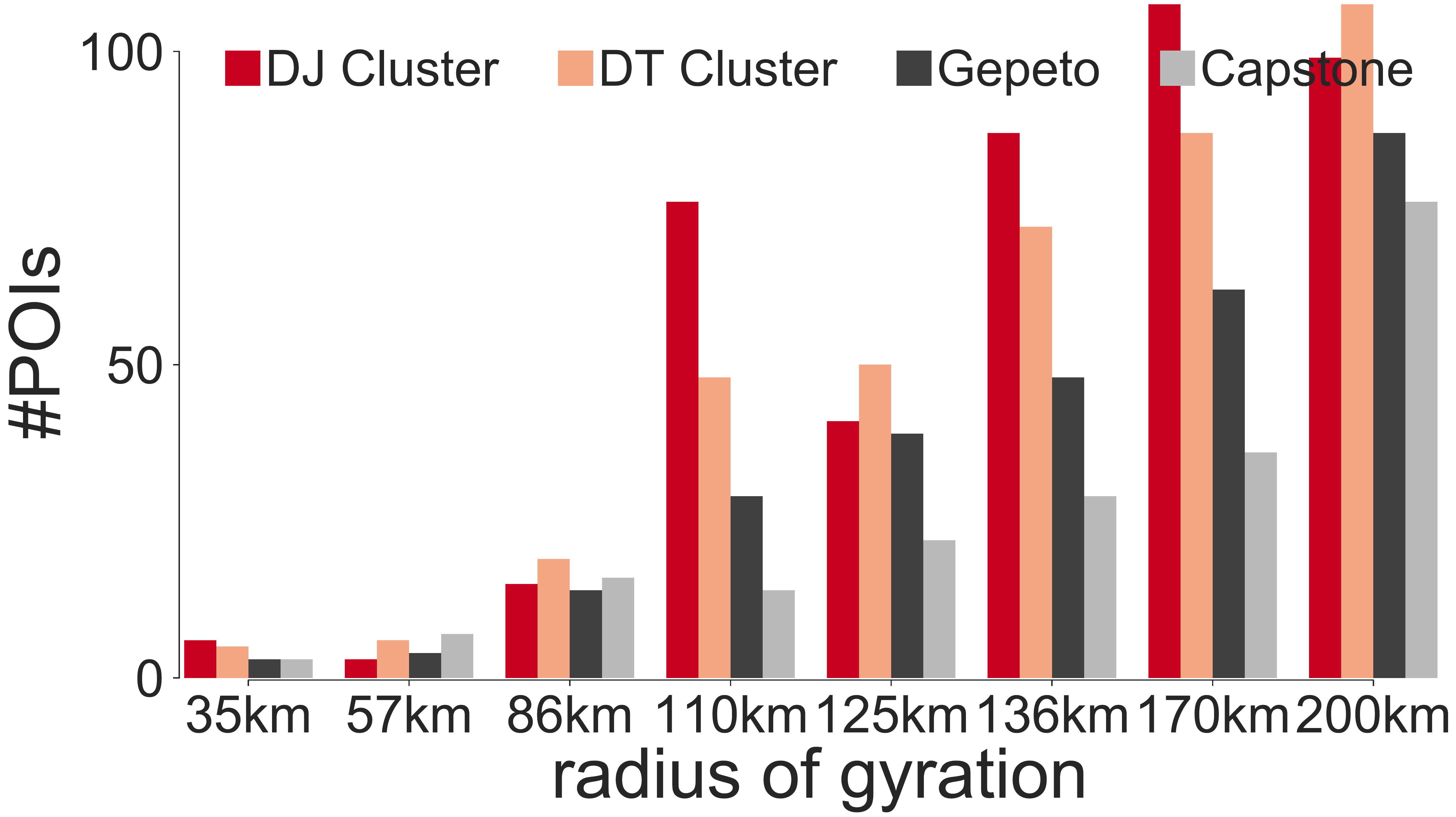}
  \captionof{figure}{Geolife Dataset}
  \label{fig:geolife_benchmark}
\end{minipage}
\vspace{-10pt}
\end{figure}

The socio-demographic information collected from the participants include age group, employment/marital status, mode of transportation during the week/week-end, field of study, level of education, frequency of participation in sport activities, allergies, smoking habits among several others. 
In order to obtain the ground-truth regarding the points of interest, we first computed all the spatiotemporal clusters present in an individual's trajectories throughout the data collection period. 
In order to select the appropriate technique, we first compared several spatiotemporal clustering approaches. 
Our objective was to select the approach that computes all the relevant clusters relying on minimal a-priori parameters while minimizing duplicate/redundant zones. 
We focus on verifying that the approach does not eliminate any true positives while allowing for the presence of false positives under a satisfactory threshold. 
It is crucial to have approximately equal distribution between true positives and false positives in order to train supervised machine learning approaches. 

In order to analyze the sensitivity thresholds of the clustering algorithms in extracting POISs , we benchmark the performance of commonly used spatiotemporal clustering approaches on the MDC dataset~\cite{nmdcpaper} (same region as {\it{Breadcrumbs}}) and the Geolife dataset~\cite{geolifepaper} to assess the number of clusters extracted.
In order to inform the sensitivity analysis, we leverage a large number of existing works that have benchmarked the mobility behavior of individuals in these datasets~\cite{thomason2016identifying}.
We compared the approaches specified by Gambs et al.~\cite{gambs2010show}, (1) DJ Cluster~\cite{zhou2004discovering}, (2) DT Cluster~\cite{Chen2007DensitybasedCF}, (3) TD Cluster~\cite{gambs2010show} and a parameter-less approach (4) Capstone~\cite{Kulkarni:2017:EHW:3139958.3140002}.
In order to perform this analysis, we select users with distinct activity areas captured with respect to the radius of gyration of movement to include distinct mobility behaviors. 
As shown in Figures~\ref{fig:mdc_benchmark} and~\ref{fig:geolife_benchmark} DJ Cluster~\cite{zhou2004discovering} detects a significantly high number of POIs, not typical for an average user based on the mobility behavioral studies by Thamason et al.~\cite{thomason2016identifying}.
The parameter-less approach~\cite{Kulkarni:2017:EHW:3139958.3140002} and TD Cluster (a variant of DT-clustering) on the other hand detects fewer POIs and could potentially lead to elimination of true positives.

We therefore used a clustering approach based on DT Cluster~\cite{Chen2007DensitybasedCF} to compute all the clusters in an individual's trajectories which offers an optimal tradeoff between potentially displaying two or more clusters at a single location and omitting true POIs.
We considered that a cluster is defined by a centroid and a radius, the latter is computed using all the points contained in a cluster. 
Here, we modify the original DT-clustering approach, where we merge two clusters if they overlap with one another, by accounting for the centroids of the clusters as well as their radius.
Furthermore, we add a filter on the number of visits to a cluster and select the ones where a participant visited least~3 times during the~12 weeks of the data collection timeframe.  
These clusters were then displayed to each respective participants along with the validation option to annotate each of the clusters and further provide semantic labels using our application at the end of the collection campaign. 
The labels span 9 categories (transport, study, residency, work, sustenance, shopping, sports, leisure and other (free-text)). 
Additionally, each participant tagged their relationship with other individuals participating in the data collection campaign. 
This information along with the demographic information, geolocation points forming trajectories was aggregated upon removing all the participant identifiers to assemble the dataset. 
The preprocessing involved removal of the duplicate points and merging all the records sequentially.

\section{Quantitative Analysis}
\label{sec:quantitative_analysis}

\begin{figure}[t!]
\centering
\includegraphics[scale=0.098]{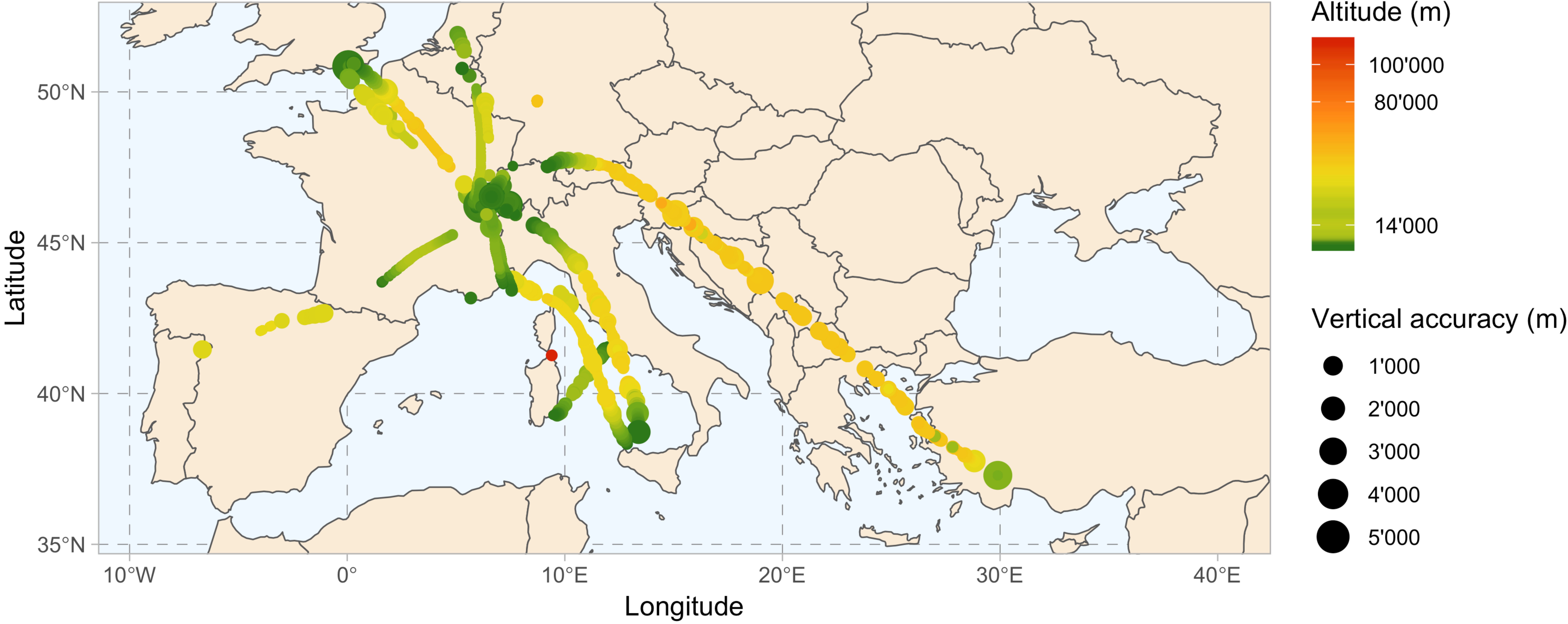}
\caption{Spatial extent of the geolocation data accompanied with the respective vertical accuracy of data-points.}
\label{fig:spatial_extent}
\vspace{-15px}
\end{figure}

In this Section we perform quantitative analysis of the {\it{Breadcrumbs}} data and present the different feature sets along with the descriptive statistics. 
{\it{Breadcrumbs}} dataset contains 46,380,042 records gathering GPS, WiFi, Bluetooth and accelerometer data points.
The dataset was collected for a period of~12 weeks from March-May 2018, retrieving geolocation data of~81 individuals at a sampling rate of~60 sec.
The aggregate distance traversed by the participants amount to 548,210 km and the average distance covered per participant is approximately 6,768 km.
Figure~\ref{fig:spatial_extent} shows the spatial extent traversed by the participants.
The largest age groups present in the campaign were 18-21 and 22-27, with 53\% and 44\% of the sample falling to these age ranges respectively.
57\% of the participants identified as females and 73\% participants were in a bachelors degree program and 25\% in a masters program.
The participants span the faculty of law, medicine, business, economics, literature, physics, biology, chemistry and computer science.\\

\begin{table}[h!]
\centering
\resizebox{0.48\textwidth}{!}{%
\begin{tabular}{llllllll}
\hline
\textbf{Variable} & \textbf{Min} & \textbf{Q25} & \textbf{Median} & \textbf{Q75} & \textbf{Max} & \textbf{Mean} & \textbf{SD} \\ \hline
\rowcolor[HTML]{EFEFEF} 
\textbf{Longitude} & -43.285 & 6.516 & 6.588 & 6.825 & 100.761 & 6.582 & 4.461 \\
\textbf{Latitude} & -22.971 & 46.317 & 46.520 & 46.537 & 55.663 & 46.217 & 2.090 \\
\rowcolor[HTML]{EFEFEF} 
\textbf{Altitude} & -450.0 & 385.659 & 415.286 & 486.020 & 111779.612 & 463.582 & 547.385 \\
\textbf{Speed} & 0.0 & 1.40 & 9.640 & 21.460 & 295.489 & 13.428 & 16.860 \\
\rowcolor[HTML]{EFEFEF} 
\textbf{Horizontal accuracy} & 0.202 & 8.0 & 12.0 & 32.0 & 149000.0 & 70.760 & 1214.937 \\
\textbf{Vertical accuracy} & 1.875 & 3.0 & 6.0 & 10.0 & 58410.027 & 14.837 & 110.618 \\ \hline
\end{tabular}%
}
\caption{Descriptive statistics of the GPS data points}
\label{tab:discrptive_stats}
\end{table}

\begin{table}[h!]
\centering
\resizebox{0.48\textwidth}{!}{%
\begin{tabular}{llllll}
\hline
\textbf{Type} & \textbf{Total \#records} & \textbf{Min per user} & \textbf{Avg \#records} & \textbf{Max per user} & \textbf{Size} \\ \hline
\rowcolor[HTML]{EFEFEF} 
\textbf{GPS} & 14656971 & 23481 & 180950 & 476445 & 1006.4 MB \\
\textbf{WiFi} & 19363007 & 17256 & 239049 & 441629 & 443.3 MB \\
\rowcolor[HTML]{EFEFEF} 
\textbf{Bluetooth} & 60986 & 0 & 753 & 5919 & 6.3 MB \\
\textbf{Accelerometer} & 12299078 & 18534 & 151840 & 421813 & 844.5 MB \\ \hline
\end{tabular}%
}
\caption{Number of data points and ratio per participant}
\label{tab:data_points}
\end{table}

\subsection{Geolocation Data}

\begin{figure}[h!]
\centering
\includegraphics[scale=0.16]{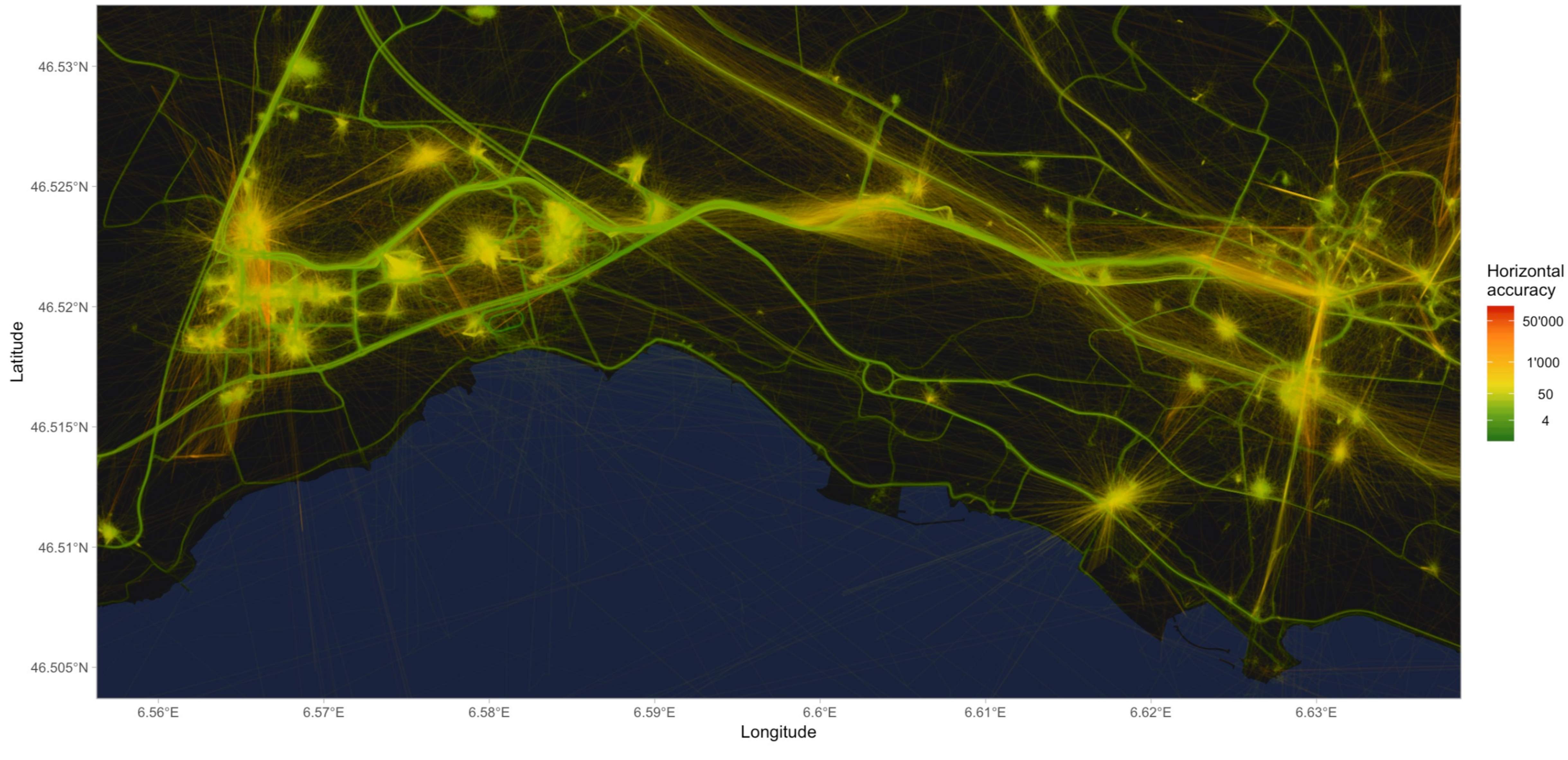}
\caption{POI clusters and horizontal accuracy of the GPS locations.}
\label{fig:poi_clusters}
\end{figure}

The summary of the GPS data points are presented in Table~\ref{tab:discrptive_stats}. 
Here, the horizontal accuracy indicates a radius about a 2-dimensional point, implying that the true unknown location is within the circle.
Vertical accuracy gives the altitude correctness of a 1-dimensional location within the region defined by the radius.
The median horizontal and vertical accuracy of the GPS data points is less than 10 meters or less.   
Table~\ref{tab:data_points} shows the number of records collected by the different sensors. 
The distribution of the POIs in the city and the horizontal accuracy of the GPS coordinates is shown in Figure~\ref{fig:poi_clusters}.\\  

\begin{figure}[h!]
\centering
\includegraphics[scale=0.3]{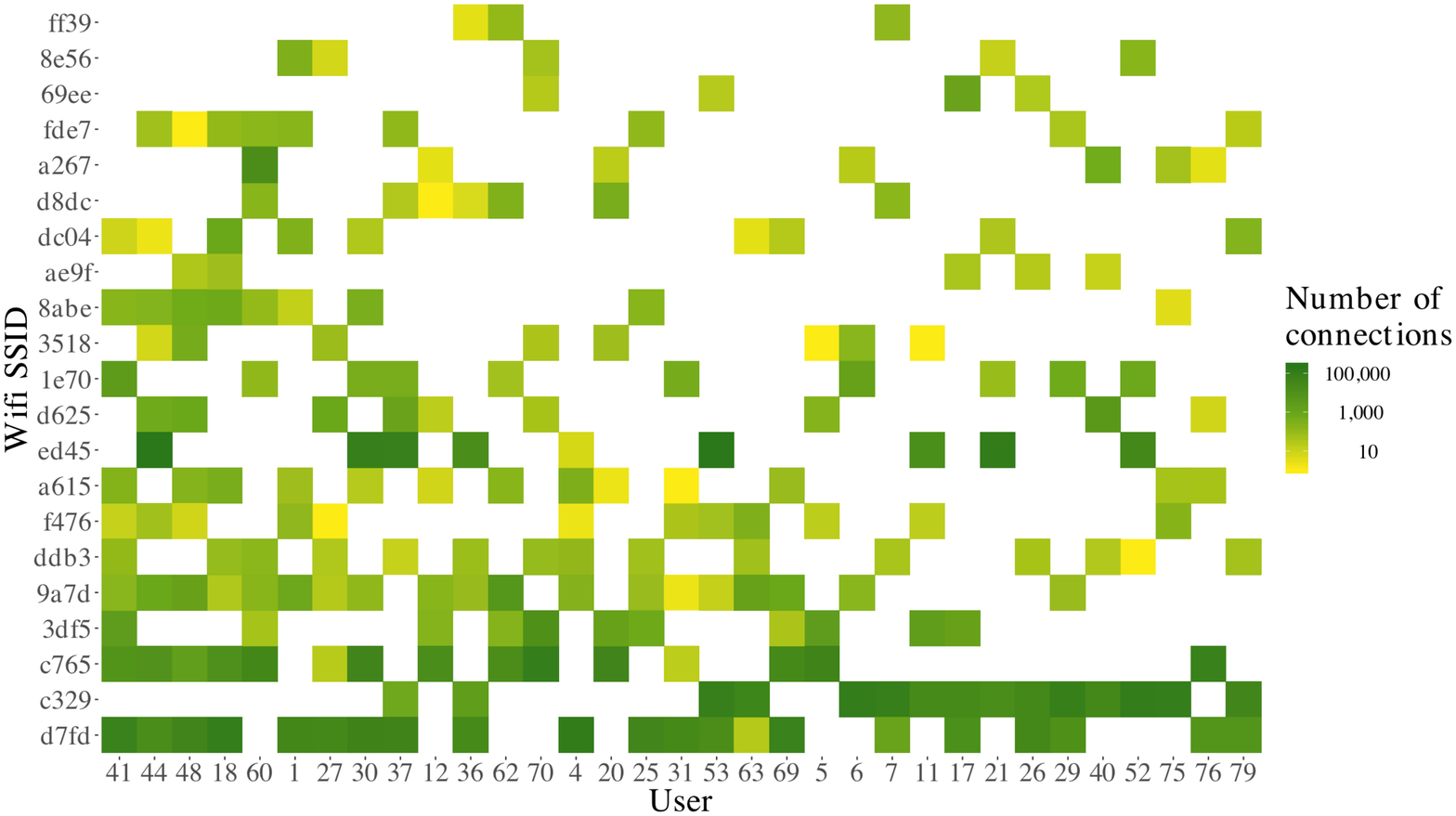}
\caption{Recurrent WiFi connections and the respective participant.}
\label{fig:recurrent_wifi}
\end{figure}

The WiFi SSIDs are not mapped to the GPS locations, instead a unique identifier corresponding to the MAC address of the WiFi access point is stored.
These identifiers act as a spatial indicator of the participant location.
The recurrent WiFi connections along with the respective participant is shown in Figure~\ref{fig:recurrent_wifi}.

\subsection{Points of Interest}

\begin{figure}[t!]
\centering
\includegraphics[scale=0.2]{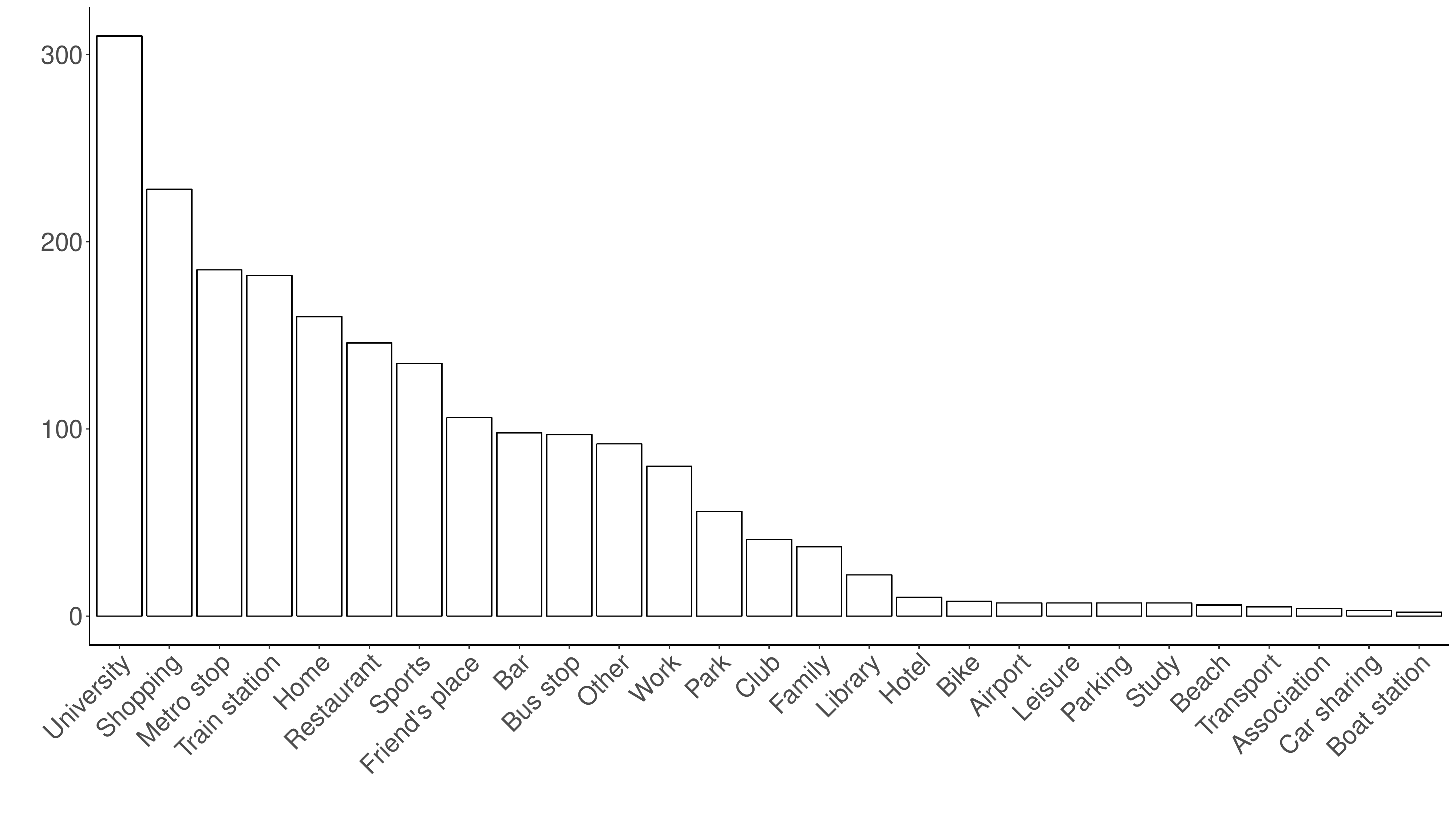}
\caption{Distribution of POIs according to their semantic labels.}
\label{fig:poi_distribution}
\end{figure}

\begin{figure}[h!]
\centering
\includegraphics[scale=0.2]{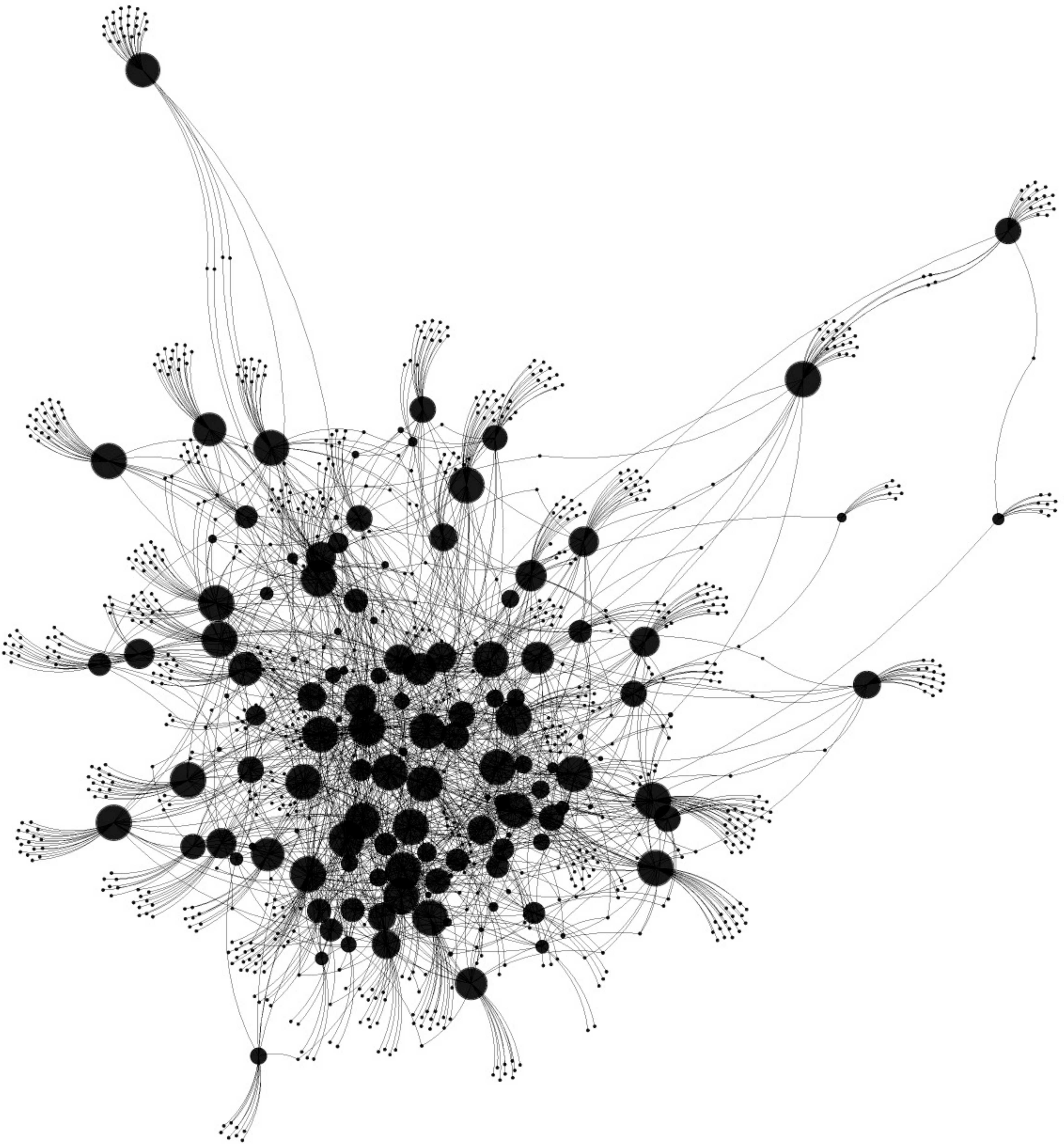}
\caption{POI to user connectivity graph.}
\label{fig:poi_connectivity}
\vspace{-10px}
\end{figure}

In this Section, we summarize the point of interest information extracted form the raw trajectories and present the descriptive statistics regarding the following aspects: (1) distribution of semantic labels, (2) connectivity graph of POI sharing, and (3) geographical overview of the POIs.
We find that a majority of the POIs are located at transport hubs, university area and leisure places as shown in Figure~\ref{fig:poi_distribution}.
We also observe the distinct POI clusters shared by a different user groups along with several isolated POIs in Figure~\ref{fig:poi_connectivity}.
A nodes in the figure denotes a unique POI and an edge signifies a user connectivity to that POI. 

\subsection{Demographic attributes}

In this section, we present the descriptive statistics pertaining to the demographics information collected using survey questionnaires. 
This information includes transportation mode preference of the participants, health related information and information such as parents' home region and high school location.

\begin{figure}[h!]
\centering
\includegraphics[scale=0.37]{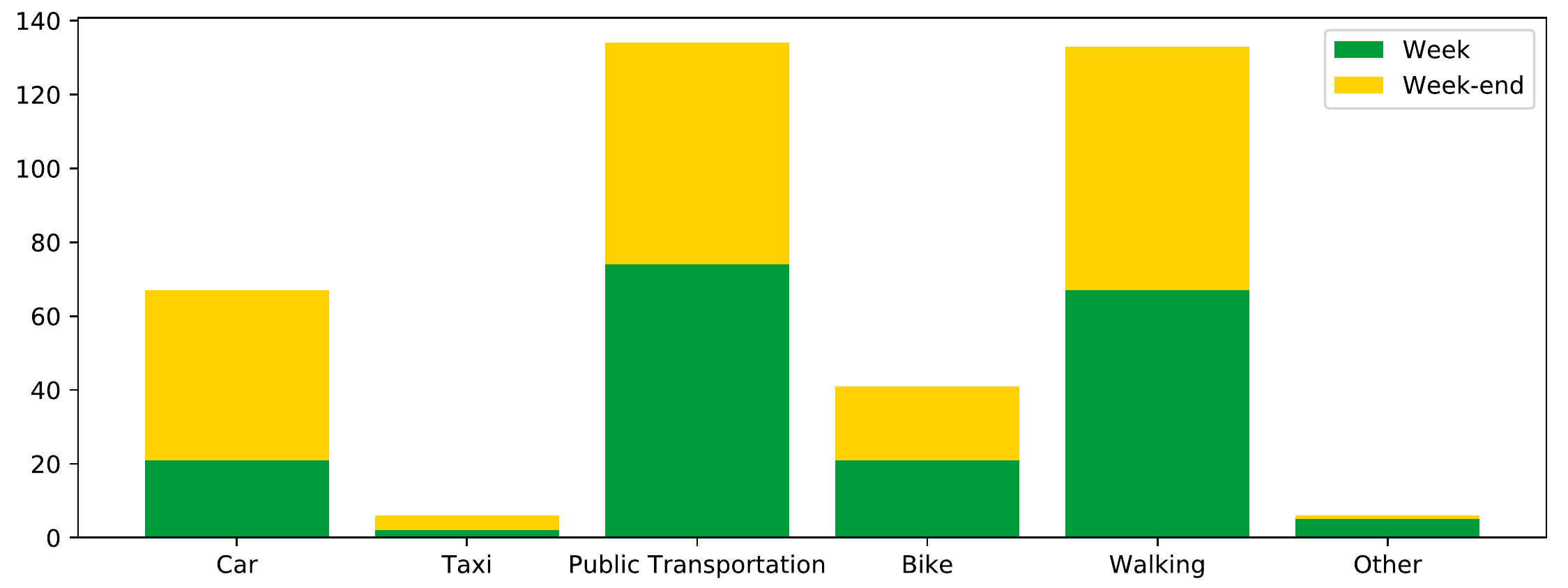}
\caption{Transportation Modes Preferences for weekday and weekend.}
\label{fig:temp_transportation_modes_preference}
\end{figure}

\begin{figure*}[h!]
\centering
\includegraphics[scale=0.62]{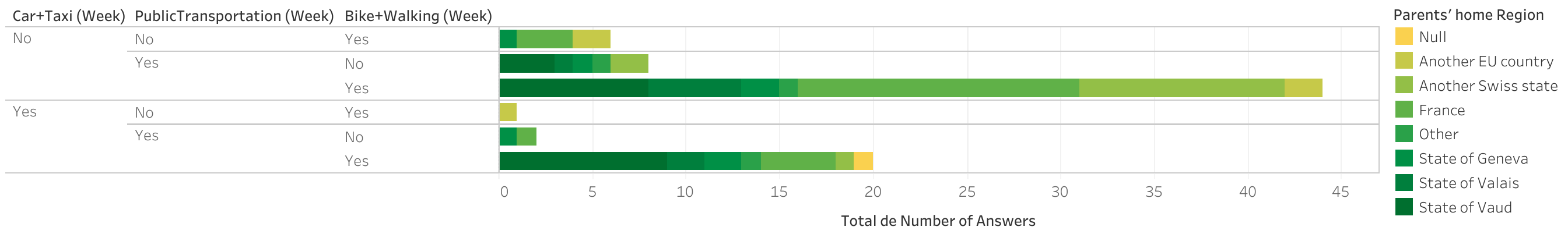}
\caption{Transportation Modes Weekly Usage And Parents' Home Region.}
\label{fig:transportation_modes_weekly_usage_parents_home_region}
\vspace{-10pt}
\end{figure*}

From this survey, we have highlighted mobility trends related to transportation modes preferences and patterns as preliminary results.
Figure~\ref{fig:temp_transportation_modes_preference} shows the transportation mode preferences during weekday and weekend.
We observe an increased usage of private transport (cars) during the week-end compared to the weekdays where the participants rely on public transportation, including trams, bus and trains.
The usage of bikes and walking preference is similar during the weekdays and the weekend.
Figure~\ref{fig:transportation_modes_weekly_usage_parents_home_region} presents the weekly mobility pattern choices according to the parents' home region.
We observe that the most represented patterns are \emph{Public Transportation + Bike/Walking} and \emph{Car + Public Transportation + Bike/Walking}.
Secondly, the figure also highlights that the least represented is the individuals who only use cars on a weekly basis.
The most represented parents' home regions of the individuals who follow the first pattern are located in \emph{France} and \emph{Another Swiss state}.
The most represented parents' home region of the individuals who follow the second pattern is located in \emph{Canton de Vaud} (this Swiss district includes the Lausanne region).
This finding indicates that most of the individuals of the second pattern study very close to their parent's home region.

\begin{figure*}[h!]
\centering
\includegraphics[scale=0.62]{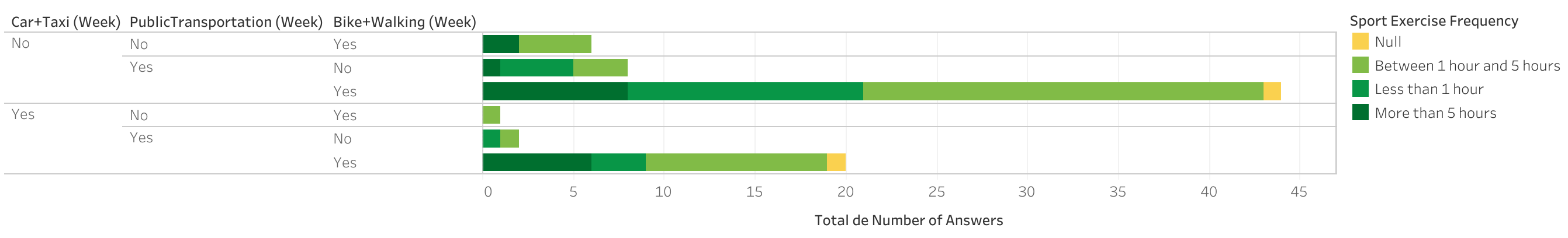}
\caption{Transportation Modes Weekly Usage And Sport Exercise Frequency.}
\label{fig:transportation_modes_weekly_usage_sport_exercise_frequency}
\vspace{-10pt}
\end{figure*}

\begin{figure*}[h!]
\centering
\includegraphics[scale=0.62]{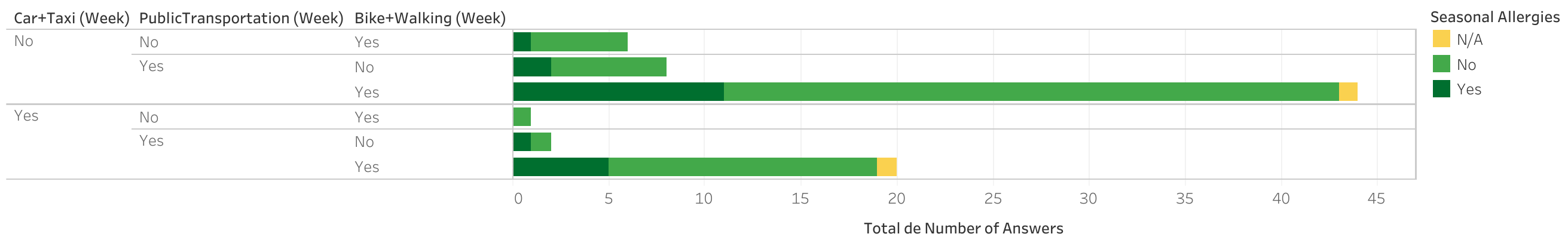}
\caption{Transportation Modes Weekly Usage And Seasonal Allergies.}
\label{fig:transportation_modes_weekly_usage_seasonal_allergies}
\vspace{-10pt}
\end{figure*}

\begin{figure*}[h!]
\centering
\includegraphics[scale=0.65]{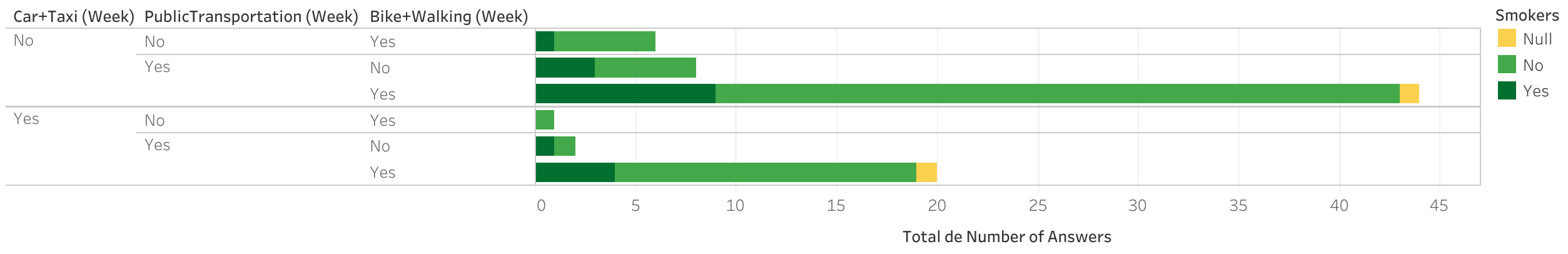}
\caption{Transportation Modes Weekly Usage And Smokers.}
\label{fig:transportation_modes_weekly_usage_smokers}
\vspace{-10pt}
\end{figure*}

\begin{figure*}[h!]
\centering
\includegraphics[scale=0.62]{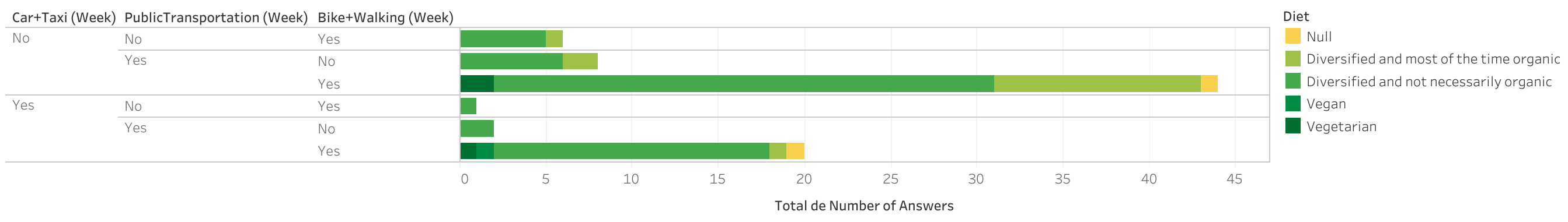}
\caption{Transportation Modes Weekly Usage And Diet.}
\label{fig:transportation_modes_weekly_usage_diet}
\vspace{-10pt}
\end{figure*}

In Figures~\ref{fig:transportation_modes_weekly_usage_sport_exercise_frequency}, \ref{fig:transportation_modes_weekly_usage_seasonal_allergies}, \ref{fig:transportation_modes_weekly_usage_smokers} and \ref{fig:transportation_modes_weekly_usage_diet}, we superimposed health characteristics on weekly transportation mode preferences.
Health characteristics are the following: frequency of sport exercise of the Breadcrumbs' participants, if they have seasonal allergies or not, if they smoke or not and the type of diet of the participants.
The figures highlight that some specific health characteristics may be related to some particular transportation modes preferences.
For example, the majority of the Breadcrumbs' participants who eat a diversified food and most of the time organic seem not use cars during the week in Figure~\ref{fig:transportation_modes_weekly_usage_diet}.

\section{Clustering Comparison \& Validation}\label{clustering_comparison}

In this section, we perform a comparative analysis of four clustering approaches that aim at extracting POIs belonging to all the participants of the {\it{Breadcrumbs}} project. 
We also focus on describing the usage of the ground-truth information to validate and compare the performance of different clustering algorithms.

\subsection{Clustering Algorithm Descriptions}

\emph{K}-means and DBSCAN only account for the spatial dimension of the locations, while DJ Cluster and DT Cluster utilize both, the spatial and temporal dimensions of the locations.

\noindent{\bfseries{k-means.}} \emph{k}-means is a widely used spatiotemporal clustering algorithm. 
Amongst the different versions of \emph{k}-means, we rely on the one proposed by Hartigan and Wong~\cite{hartigan1979algorithm}.
The algorithm takes two input values: \emph{k}, indicating the number of clusters to be obtained, and the set of data-points to be be grouped in to~\emph{k} clusters.
The output is $k$ clusters representing the user points of interest. 
The algorithm operates as follows:

\begin{itemize}

	\item \emph{k} points are randomly chosen in the initial set of points and considered as the initial centroids of the \emph{k} clusters;
	\item All the points of the initial set of points are then assigned with their closest centroid based on the Euclidean distance between points and the centroids;
	\item All the centroids of the \emph{k} clusters are then updated by calculating the mean of all points being associated to a cluster;
	\item Finally the algorithm iterates until converging into a stable state in which there is no more additional way to minimize the total sum of squared Euclidean distances between points and their related centroid.
	
\end{itemize}

\noindent{\bfseries{DBSCAN.}} DBSCAN is a clustering algorithm based on the density of the points linked to clusters.
Unlike \emph{k}-means, we do not need to specify the number clusters \emph{a priori} and is not known beforehand.
The algorithm takes three input values: \emph{eps}, which is the maximum radius of the neighborhood of a point, \emph{minPts}, which indicates the minimum number of points that must be in the neighborhood of a point, and the initial set of points (i.e., locations) that must be grouped in several clusters.
These clusters represent the user POIs. 
DBSCAN operates as follows:

\begin{itemize}

	\item The algorithm starts evaluating each data-point of the initial by computing the density with respect to all the other points using the \emph{eps} value and the \emph{minPts}. This step associates a category to each point between the core, border and noise points. Noise points are later deleted;
	\item Neighborhood core points, which are density-reachable, are then associated with the same cluster;
	\item Finally, border points are associated to the nearest cluster.
	
\end{itemize}

\noindent{\bfseries{DJ Cluster.}} DJ Cluster is a clustering algorithm that takes into account spatiotemporal dimensions~\cite{zhou2004discovering}.
The algorithm takes four input parameters: \emph{minPts}, which is minimum of points located in a candidate cluster, \emph{r}, which is the radius of a candidate cluster, \emph{speed-threshold}, which is maximum speed we will use to extract the meaningful points from which the candidate clusters are built, and the initial set of points (i.e., locations).
\begin{itemize}

	\item The algorithm initiates by deleting all moving points (a moving point is deleted if the time difference with the previous point is greater than \emph{speed-threshold});
	\item Then, we compute the neighborhood density of each remaining point and only keep the points that have at least \emph{minPts} points in a specific \emph{r}. The remaining points of this step are the candidate clusters;
	\item Finally, a merging step is realized. The algorithm merges the candidates clusters that share at least one common point.
	
\end{itemize}

\noindent{\bfseries{DT Cluster.}} DT Cluster is a spatiotemporal clustering algorithm~\cite{Chen2007DensitybasedCF}.
The algorithm takes three input parameters: \emph{d}, which is the radius of the candidate clusters created at the beginning of the process, \emph{t}, which is minimum time period spent in the candidate clusters also created at the beginning of the algorithm, and the initial set of points (i.e., locations).

\begin{itemize}

	\item The algorithm starts by evaluating the full set of points (i.e., locations) like a time series. It extracts all candidate clusters by gathering the successive points that are located in the same \emph{d} for a duration greater than \emph{t};
	\item To finish the process, a merging step is performed. This step merges all the candidate clusters that are lie at a distance \emph{d/3} from each other.

\end{itemize}

\subsection{Evaluation Framework and Parameters}

\begin{figure}[h!]
\centering
\includegraphics[scale=0.5]{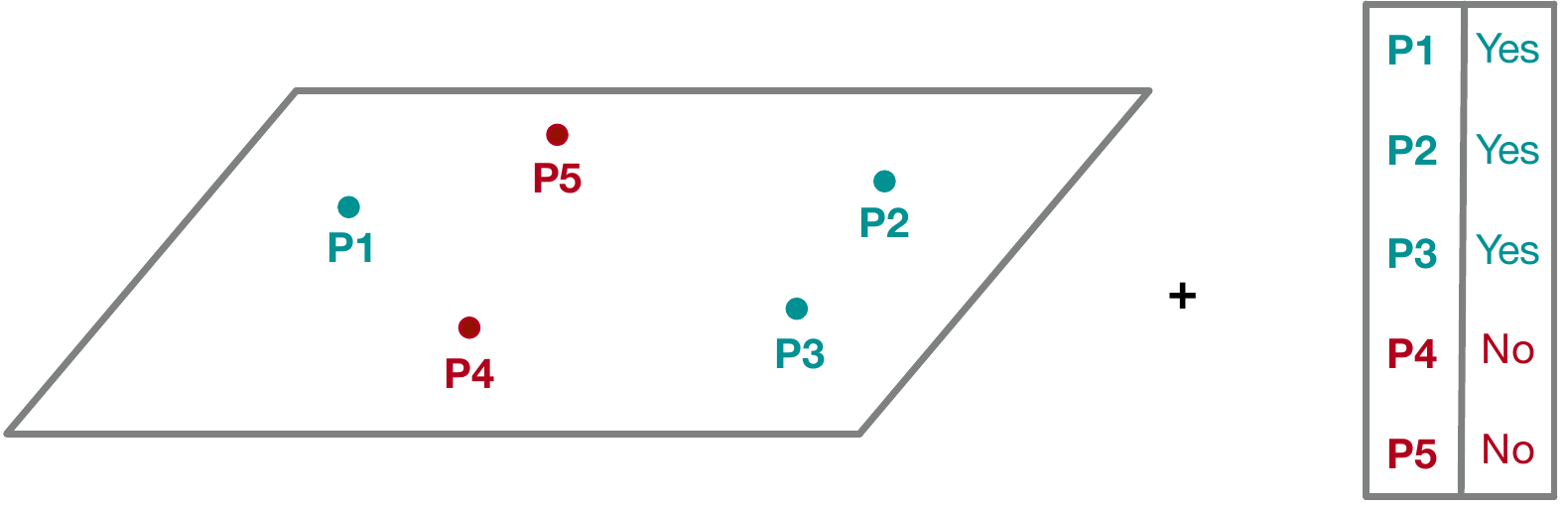}
\caption{Ground-truth and annotation (validation by the participant with a yes or a no for each point detected)}
\label{fig:evaluation_procedure_step_1}
\vspace{-10pt}
\end{figure}

In this section, we present the performance evaluation framework.
The evaluation is based on two steps that will be described hereafter.
For sake of simplicity, we will take the example of one single {\it{Breadcrumbs}} participant.
However, the evaluation in the results section considers all the 81 participants. 
The evaluation takes as input the ground-truth information labelled by the user and a clustering algorithm
Figure~\ref{fig:evaluation_procedure_step_1} presents the ground-truth of the participant as well as the annotations (i.e., yes or no) linked to every points.
The valid POIs are therefore the points labelled as {\it{yes}} by the participant.
The points labeled {\it{no}} are as well crucial to the evaluation framework. 

\begin{figure}[h!]
\centering
\includegraphics[scale=0.49]{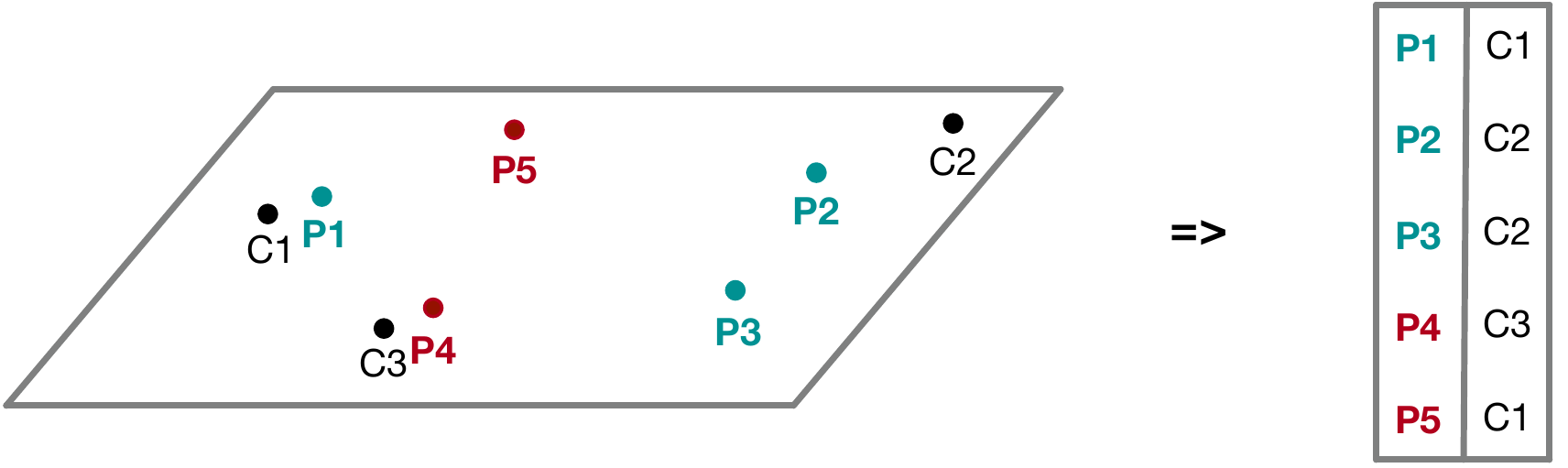}
\caption{Links between Ground-truth (P) and Clusters (C) according to a minimum distance}
\label{fig:evaluation_procedure_step_2}
\vspace{-10pt}
\end{figure}

The first step consists in linking each point of the ground-truth with a corresponding cluster, which was extracted by the clustering algorithm.
To do so, we will find the closest cluster to each point of the ground-truth by computing the euclidian distance between them.
The output of step one is illustrated in Figure~\ref{fig:evaluation_procedure_step_2}.

\begin{figure}[h!]
\centering
\includegraphics[scale=0.36]{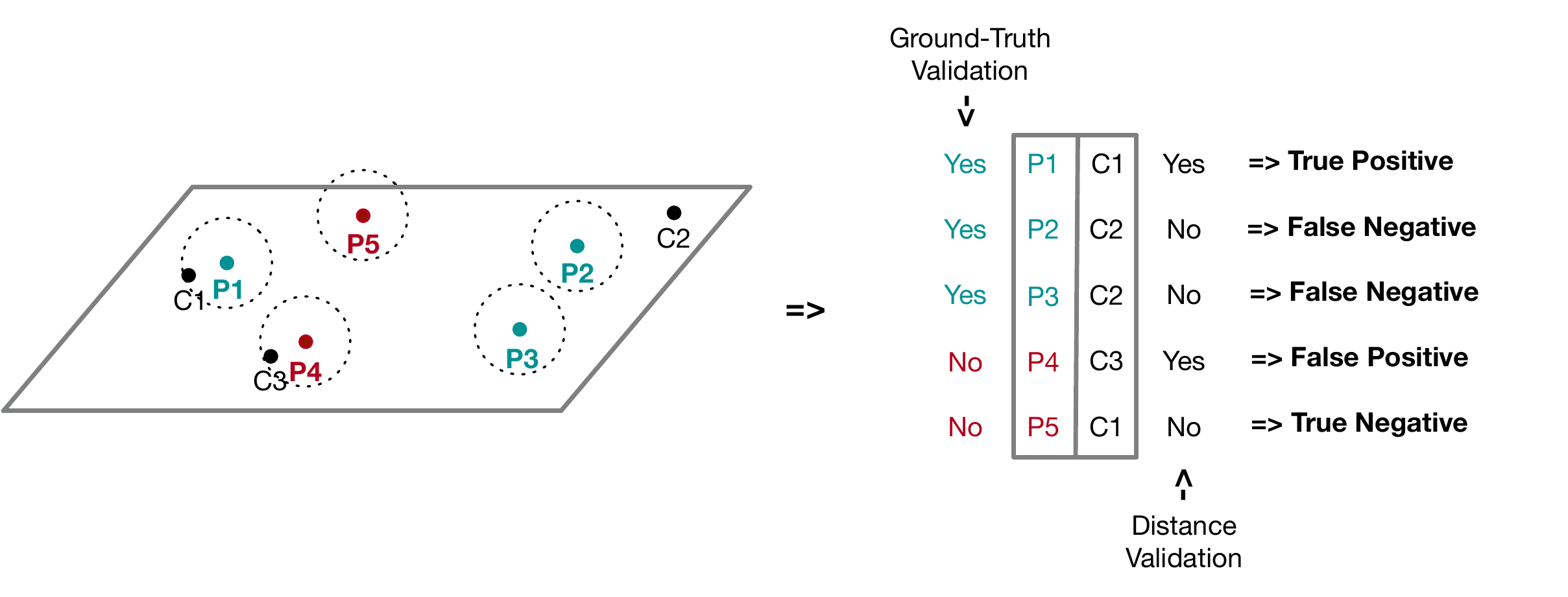}
\caption{Final Annotation (TP, TN, FP and FN)}
\label{fig:evaluation_procedure_step_3}
\vspace{-10pt}
\end{figure}

The second step aims at extracting the number of true positives, false positives, true negatives and false negatives that will enable us to compute the true positive rate (sensitivity) and false positive rate (1 - specificity) in order to formulate the ROC (Receiver Operating Characteristic) curve.
In order to perform this step, we will use a parameter~\emph{d} that will help to determine if a cluster is located in a validation zone around a point of the ground-truth.
The validation results are described in the column \emph{Distance Validation} in Figure~\ref{fig:evaluation_procedure_step_3}.
The true positives (TP) have a ground-truth validation and a distance validation that correspond to {\it{Yes}}, while the true negatives (TN) have a ground-truth validation and a distance validation that are equal to {\it{No}}.
The false positives have a ground-truth validation value that is equal to {\it{No}} and a distance validation value to {\it{Yes}}.
Similarly the false negatives posses contrary labels to the above. 

The ROC curve is computed with the true positive rate, indicated on the y-axis, and true negative rate, indicated on the x-axis.

\begin{itemize}

	\item True positive rate (sensitivity): True Positives / (True Positives + False Negatives)
	\item False positive rate (1 - specificity): False Positives / (False Positives + True Negatives)
	
\end{itemize}

\begin{table}[t!]
\centering
\resizebox{0.47\textwidth}{!}{%
\begin{tabular}{llllllll}
\hline
\textbf{Clustering Algorithm} & \textbf{Parameter Values} \\ \hline
\rowcolor[HTML]{EFEFEF} 
\textbf{\emph{k}-means} & k = 10/30/100/200/300/1000 \\
\textbf{DBSCAN} & minPts = 30 / eps = 0.003/0.001/0.0007/0.0002/0.0001 \\
\rowcolor[HTML]{EFEFEF} 
\textbf{DJ Cluster} & radius = 60.0 m / speed-threshold = 1.5 km/h / minPts = 10/20/50/100/200/500 \\
\textbf{DT Cluster} & t = 15 mins (900 sec) / d = 40/60/100/150/300 m  \\ \hline
\end{tabular}%
}
\caption{Parameters of the Clustering Algorithms}
\label{tab:clustering_algorithm_parameters}
\end{table}

\begin{figure}[h!]
\centering
\includegraphics[scale=0.38]{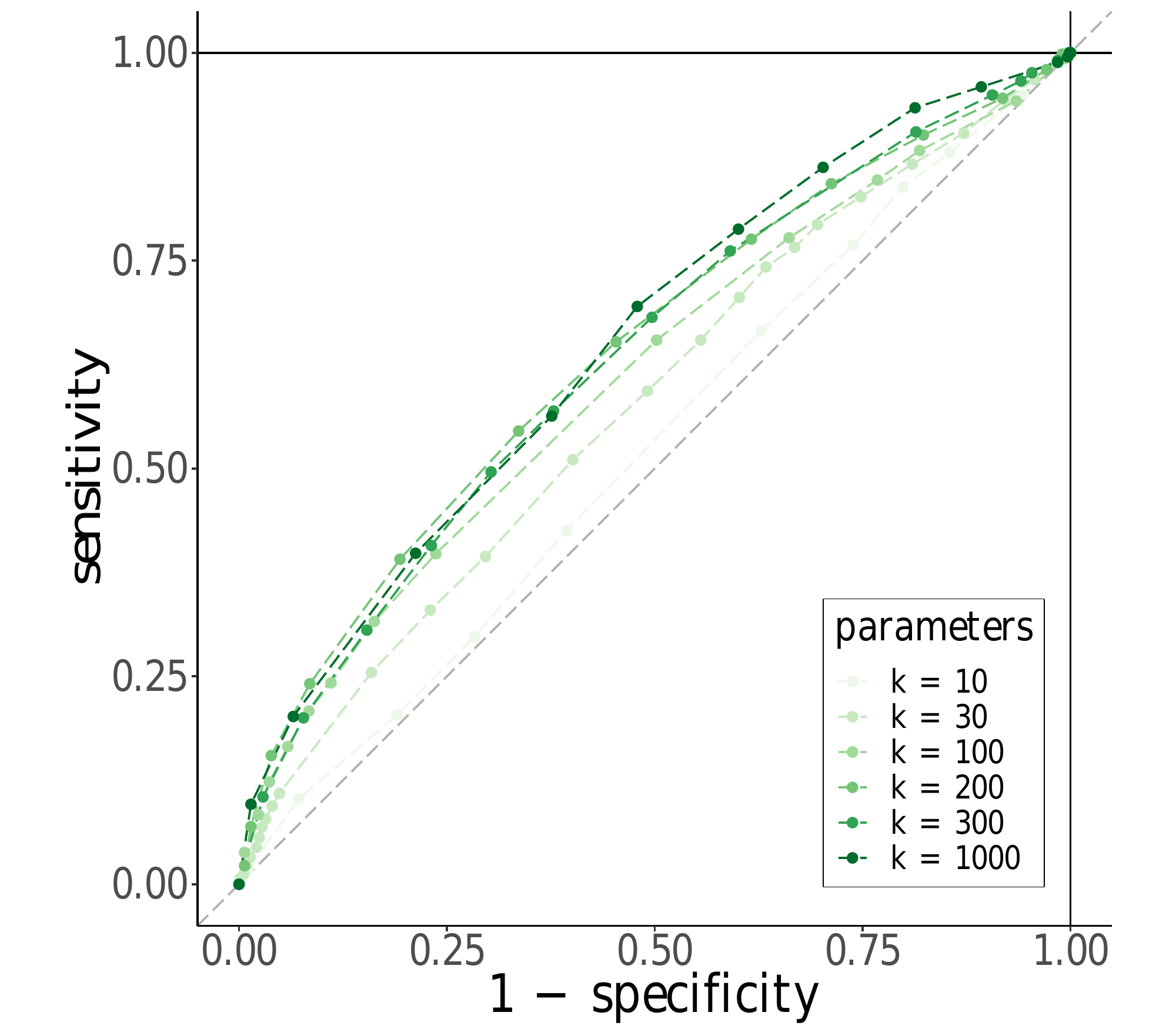}
\caption{ROC Curve - \emph{k}-means Clustering Algorithm}
\label{fig:roc_k_mean}
\vspace{-10pt}
\end{figure}

\begin{figure}[h!]
\centering
\includegraphics[scale=0.38]{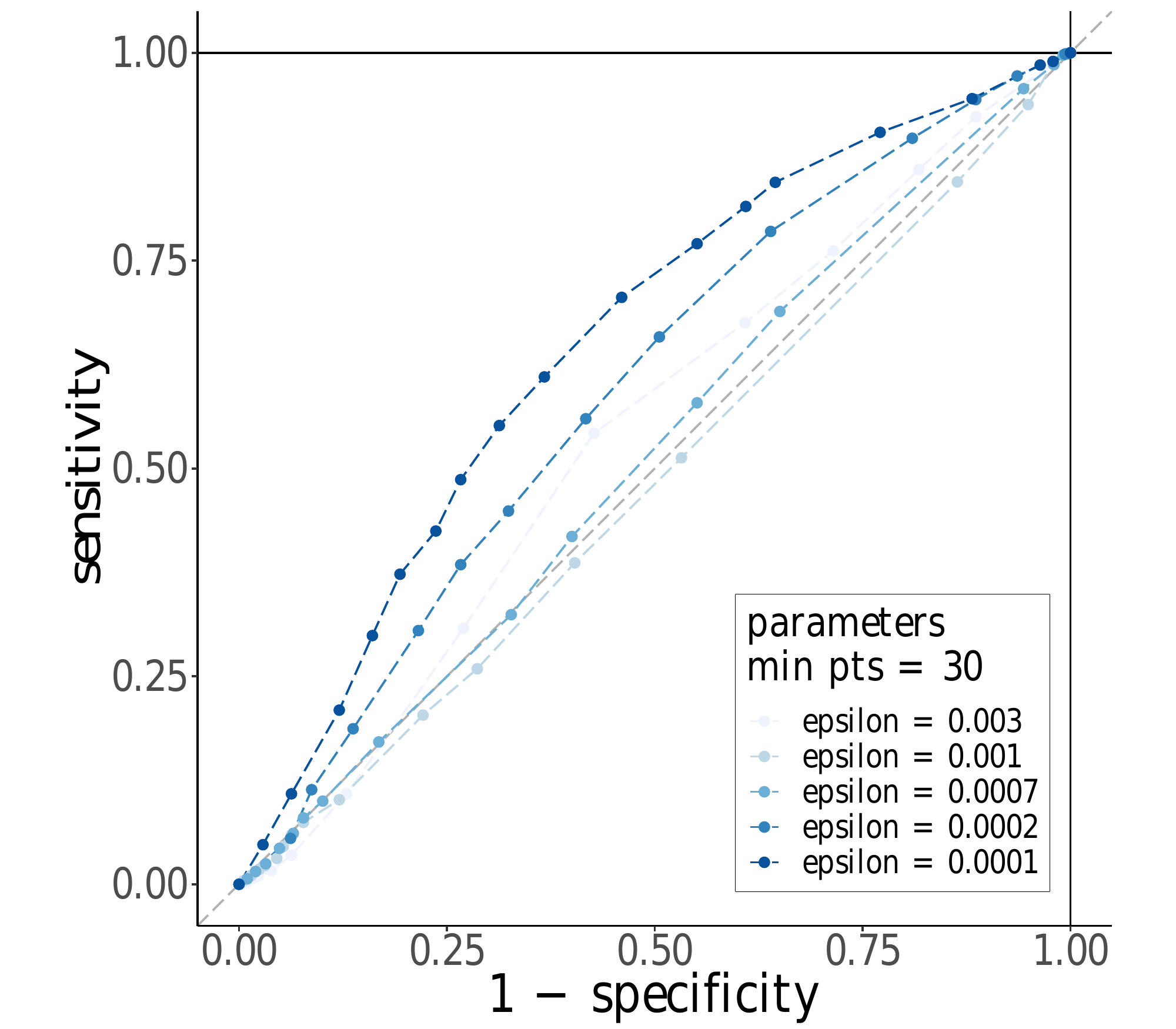}
\caption{ROC Curve - DBSCAN Clustering Algorithm}
\label{fig:roc_db_scan}
\vspace{-10pt}
\end{figure}

\begin{figure}[h!]
\centering
\includegraphics[scale=0.38]{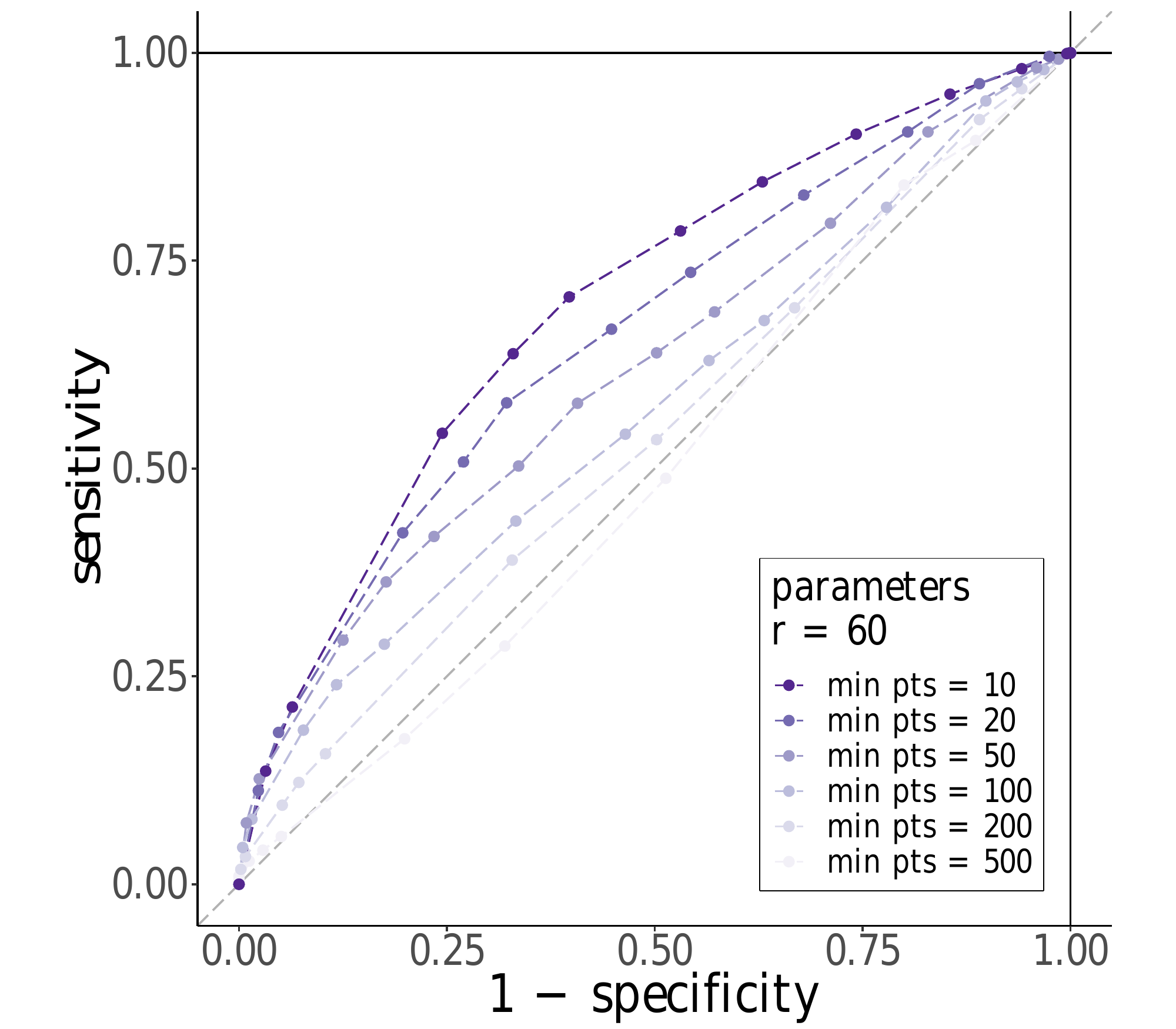}
\caption{ROC Curve - DJ Cluster Algorithm}
\label{fig:roc_dj}
\vspace{-10pt}
\end{figure}

\begin{figure}[h!]
\centering
\includegraphics[scale=0.38]{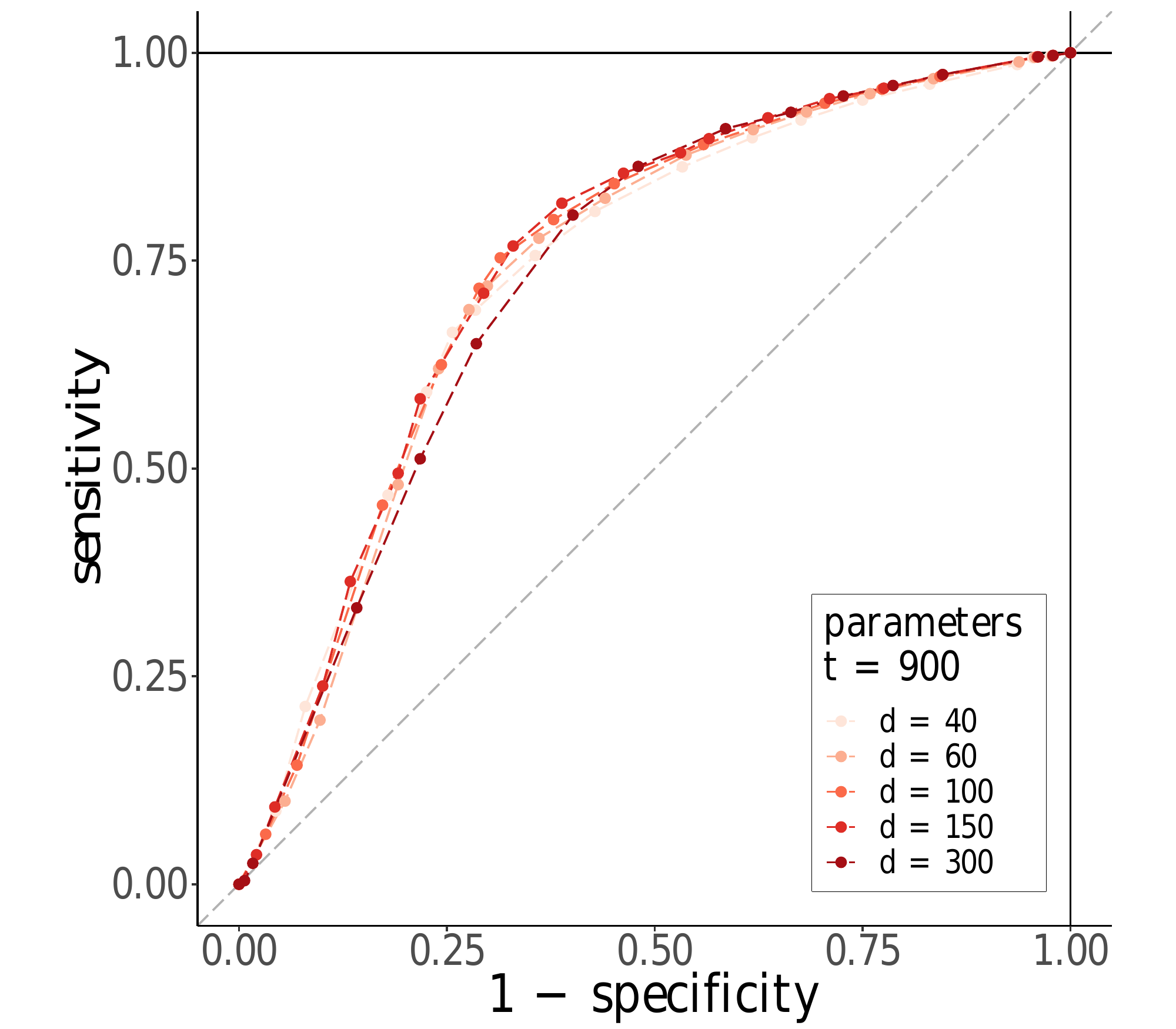}
\caption{ROC Curve - DT Cluster Algorithm}
\label{fig:roc_dt}
\vspace{-10pt}
\end{figure}

Table~\ref{tab:clustering_algorithm_parameters} describes all the selected parameters for each of clustering algorithms.
We selected the parameters according to previous research works in the same area~\cite{moro2017,kulkarni:2016:MMP:3003421.3003424} and the more plausible and values according to the spatial or spatiotemporal clustering algorithm context.

\subsection{Results}

The results of our comparative analysis are shown in Figures~\ref{fig:roc_k_mean}, \ref{fig:roc_db_scan}, \ref{fig:roc_dj} and \ref{fig:roc_dt} for \emph{k}-means, DBSCAN, DJ Cluster and DT Cluster respectively.
It is crucial to notice that the diagonal (the light gray dotted line from the bottom left corner to the top right corner) on each graph represents the worst case situation in which the algorithm has no discrimination capability to identify a cluster as a POI.
Regarding \emph{k}-means in Figure~\ref{fig:roc_k_mean}, increasing the \emph{k} marginally increases the performance until reaching a limit whereafter the performance gain halts.  
As observed in Figure~\ref{fig:roc_db_scan}, DBSCAN depicts a better performance as compared to \emph{k}-means.
DBSCAN introduces the notion of density of the neighborhood of each point evaluated, which increases the accuracy compared to the \emph{k}-means 
The lower the epsilon (that indicates the area evaluated around a point), the higher is the algorithmic performance.
We also observe that DJ cluster performs better as compared to both DBSCAN and \emph{k}-means (see Figure~\ref{fig:roc_dj}).
Furthermore, the parameter \emph{minPts} when set to the least number provides the best results
This can be justified based on the lower values of the radius \emph{r} (i.e., 60 meters) in addition to deletion of all the moving points. 
DT Cluster provides the best overall performance as observed in Figure~\ref{fig:roc_dt}
We also notice that the highest value of parameter~\emph{d} (i.e., 300 meters) is not necessarily correlated with the best performance 
Based on these results we can conclude that the clustering approaches account for the spatiotemporal parameters are comparatively better adapt to extract POIs.

\section{Conclusion}\label{sec:conclusion}

In this paper, we introduce a feature rich geolocation mobility dataset {\it{Breadcrumbs}}.
In addition to fine-grained demographic attributes, contact and calendar records, social relationships we also provide ground truth and semantic labels for the points of interest. 
We describe the complete data collection process and our methodology to collect ground-truth information in GIS domain.
Our qualitative analysis has shed light on several aspects of this dataset including the POI connectivity, WiFI connection recurrence, POI distribution.
We specify the use cases, applicable research domains and validation methodologies using the unique features of {\it{Breadcrumbs}}.
In addition, we have also highlighted the utility of health related information and transportation modes preferences.
To showcase a use case of our dataset, we have performed a comparative study of four clustering approaches to extract POI cluster from GPS trajectories. 
We have proposed a validation methodology while using the ground-truth labels illustrated the obtained results.   
We learn that DT Clustering outperforms DJ cluster, DBSCAN and \emph{k}-means and we discuss its implication.
We make {\it{Breadcrumbs}} accessible to the research community in order to facilitate and advance the GIS research.

{\scriptsize{
\bibliographystyle{abbrv}
\bibliography{sig-alternate-sample}

\begin{thebibliography}{10}

\bibitem{Ashbrook2003UsingGT}
D.~Ashbrook and T.~Starner.
\newblock Using gps to learn significant locations and predict movement across
  multiple users.
\newblock {\em Personal and Ubiquitous Computing}, 7:275--286, 2003.

\bibitem{backes2017walk2friends}
M.~Backes, M.~Humbert, J.~Pang, and Y.~Zhang.
\newblock walk2friends: Inferring social links from mobility profiles.
\newblock In {\em Proceedings of the 2017 ACM SIGSAC Conference on Computer and
  Communications Security}, pages 1943--1957. ACM, 2017.

\bibitem{barabasi2005origin}
A.-L. Barabasi.
\newblock The origin of bursts and heavy tails in human dynamics.
\newblock {\em Nature}, 435(7039):207, 2005.

\bibitem{bell2011nodobo}
S.~Bell, A.~McDiarmid, and J.~Irvine.
\newblock Nodobo: Mobile phone as a software sensor for social network
  research.
\newblock In {\em 2011 IEEE 73rd vehicular technology conference (VTC Spring)},
  pages 1--5. IEEE, 2011.

\bibitem{bogomolov2014once}
A.~Bogomolov, B.~Lepri, J.~Staiano, N.~Oliver, F.~Pianesi, and A.~Pentland.
\newblock Once upon a crime: towards crime prediction from demographics and
  mobile data.
\newblock In {\em Proceedings of the 16th international conference on
  multimodal interaction}, pages 427--434. ACM, 2014.

\bibitem{Calabrese2011EstimatingOF}
F.~Calabrese, G.~D. Lorenzo, L.~Liu, and C.~Ratti.
\newblock Estimating origin-destination flows using mobile phone location data.
\newblock {\em IEEE Pervasive Computing}, 10:36--44, 2011.

\bibitem{campbell2008rise}
A.~T. Campbell, S.~B. Eisenman, N.~D. Lane, E.~Miluzzo, R.~A. Peterson, H.~Lu,
  X.~Zheng, M.~Musolesi, G.-S. Ahn, et~al.
\newblock The rise of people-centric sensing.
\newblock {\em IEEE Internet Computing}, (4):12--21, 2008.

\bibitem{Chakka2003IndexingLT}
V.~P. Chakka, A.~Everspaugh, and J.~M. Patel.
\newblock Indexing large trajectory data sets with seti.
\newblock In {\em CIDR}, 2003.

\bibitem{chapuis2018geodabs}
B.~Chapuis and B.~Garbinato.
\newblock Geodabs: Trajectory indexing meets fingerprinting at scale.
\newblock In {\em 2018 IEEE 38th International Conference on Distributed
  Computing Systems (ICDCS)}, pages 1086--1095. IEEE, 2018.

\bibitem{chapuis2016capturing}
B.~Chapuis, A.~Moro, V.~Kulkarni, and B.~Garbinato.
\newblock Capturing complex behaviour for predicting distant future
  trajectories.
\newblock In {\em Proceedings of the 5th ACM SIGSPATIAL International Workshop
  on Mobile Geographic Information Systems}, pages 64--73. ACM, 2016.

\bibitem{Chen2007DensitybasedCF}
Y.~Chen and L.~Tu.
\newblock Density-based clustering for real-time stream data.
\newblock In {\em KDD}, 2007.

\bibitem{chessa2017mobile}
S.~Chessa, M.~Girolami, L.~Foschini, R.~Ianniello, A.~Corradi, and
  P.~Bellavista.
\newblock Mobile crowd sensing management with the participact living lab.
\newblock {\em Pervasive and Mobile Computing}, 38:200--214, 2017.

\bibitem{cho2011friendship}
E.~Cho, S.~A. Myers, and J.~Leskovec.
\newblock Friendship and mobility: user movement in location-based social
  networks.
\newblock In {\em Proceedings of the 17th ACM SIGKDD international conference
  on Knowledge discovery and data mining}, pages 1082--1090. ACM, 2011.

\bibitem{Cho2011FriendshipAM}
E.~Cho, S.~A. Myers, and J.~Leskovec.
\newblock Friendship and mobility: user movement in location-based social
  networks.
\newblock In {\em KDD}, 2011.

\bibitem{upb-hyccups-20161017}
R.~I. Ciobanu and C.~Dobre.
\newblock {CRAWDAD} dataset upb/hyccups (v. 2016-10-17).
\newblock Downloaded from \url{https://crawdad.org/upb/hyccups/20161017}, Oct.
  2016.

\bibitem{CronenTownsend2002PredictingQP}
S.~Cronen-Townsend, Y.~Zhou, and W.~B. Croft.
\newblock Predicting query performance.
\newblock In {\em SIGIR}, 2002.

\bibitem{duntgen2009berlinmod}
C.~D{\"u}ntgen, T.~Behr, and R.~H. G{\"u}ting.
\newblock Berlinmod: a benchmark for moving object databases.
\newblock {\em The VLDB Journal—The International Journal on Very Large Data
  Bases}, 18(6):1335--1368, 2009.

\bibitem{Eagle2005RealityMS}
N.~Eagle and A.~Pentland.
\newblock Reality mining: sensing complex social systems.
\newblock {\em Personal and Ubiquitous Computing}, 10:255--268, 2005.

\bibitem{elsalamouny2016differential}
E.~ElSalamouny and S.~Gambs.
\newblock Differential privacy models for location-based services.
\newblock {\em Transactions on Data Privacy}, 9(1):15--48, 2016.

\bibitem{etter2012been}
V.~Etter, M.~Kafsi, and E.~Kazemi.
\newblock Been there, done that: What your mobility traces reveal about your
  behavior.
\newblock Technical report, 2012.

\bibitem{FILLEKES2019193}
M.~P. Fillekes, C.~R{\"o}cke, M.~Katana, and R.~Weibel.
\newblock Self-reported versus gps-derived indicators of daily mobility in a
  sample of healthy older adults.
\newblock {\em Social Science and Medicine}, 220:193 -- 202, 2019.

\bibitem{furletti2017discovering}
B.~Furletti, R.~Trasarti, P.~Cintia, and L.~Gabrielli.
\newblock Discovering and understanding city events with big data: the case of
  rome.
\newblock {\em Information}, 8(3):74, 2017.

\bibitem{gambs2010show}
S.~Gambs, M.-O. Killijian, and M.~N. del Prado~Cortez.
\newblock Show me how you move and i will tell you who you are.
\newblock In {\em Proceedings of the 3rd ACM SIGSPATIAL International Workshop
  on Security and Privacy in GIS and LBS}, pages 34--41. ACM, 2010.

\bibitem{gambs2012next}
S.~Gambs, M.-O. Killijian, and M.~N. del Prado~Cortez.
\newblock Next place prediction using mobility markov chains.
\newblock In {\em Proceedings of the First Workshop on Measurement, Privacy,
  and Mobility}, page~3. ACM, 2012.

\bibitem{hariharan2004project}
R.~Hariharan and K.~Toyama.
\newblock Project lachesis: parsing and modeling location histories.
\newblock In {\em International Conference on Geographic Information Science},
  pages 106--124. Springer, 2004.

\bibitem{hartigan1979algorithm}
J.~A. Hartigan and M.~A. Wong.
\newblock Algorithm as 136: A k-means clustering algorithm.
\newblock {\em Journal of the Royal Statistical Society. Series C (Applied
  Statistics)}, 28(1):100--108, 1979.

\bibitem{nmdcpaper}
N.~Kiukkonen, J.~Blom, O.~Dousse, D.~Gatica-Perez, and J.~K. Laurila.
\newblock Towards rich mobile phone datasets: Lausanne data collection
  campaign.
\newblock 2010.

\bibitem{kulkarni2019examining}
V.~Kulkarni, A.~Mahalunkar, B.~Garbinato, and J.~D. Kelleher.
\newblock Examining the limits of predictability of human mobility.
\newblock {\em Entropy}, 21(4):432, 2019.

\bibitem{Kulkarni:2017:EHW:3139958.3140002}
V.~Kulkarni, A.~Moro, B.~Chapuis, and B.~Garbinato.
\newblock Extracting hotspots without a-priori by enabling signal processing
  over geospatial data.
\newblock In {\em Proceedings of the 25th ACM SIGSPATIAL International
  Conference on Advances in Geographic Information Systems}, SIGSPATIAL'17,
  pages 79:1--79:4, New York, NY, USA, 2017. ACM.

\bibitem{kulkarni:2016:MMP:3003421.3003424}
V.~Kulkarni, A.~Moro, and B.~Garbinato.
\newblock Mobidict: A mobility prediction system leveraging realtime location
  data streams.
\newblock In {\em Proceedings of the 7th ACM SIGSPATIAL International Workshop
  on GeoStreaming}, IWGS '16, pages 8:1--8:10, New York, NY, USA, 2016. ACM.

\bibitem{Leal2016TowardsAE}
E.~Leal, L.~Gruenwald, J.~Zhang, and S.~You.
\newblock Towards an efficient top-k trajectory similarity query processing
  algorithm for big trajectory data on gpgpus.
\newblock {\em 2016 IEEE International Congress on Big Data (BigData
  Congress)}, pages 206--213, 2016.

\bibitem{Li2010MiningPB}
Z.~Li, B.~Ding, J.~Han, R.~Kays, and P.~Nye.
\newblock Mining periodic behaviors for moving objects.
\newblock In {\em KDD}, 2010.

\bibitem{privamovpaper}
S.~B. Mokhtar, A.~Boutet, L.~Bouzouina, P.~Bonnel, O.~Brette, L.~Brunie,
  M.~Cunche, S.~D. 'Alu, V.~Primault, P.~Raveneau, H.~Rivano, and R.~Stanica.
\newblock Priva'mov: Analysing human mobility through multi-sensor datasets.
\newblock 2017.

\bibitem{montoliu2010discovering}
R.~Montoliu and D.~Gatica-Perez.
\newblock Discovering human places of interest from multimodal mobile phone
  data.
\newblock In {\em Proceedings of the 9th International Conference on Mobile and
  Ubiquitous Multimedia}, page~12. ACM, 2010.

\bibitem{moro2017}
A.~Moro and B.~Garbinato.
\newblock A location privacy estimator based on spatio-temporal location
  uncertainties.
\newblock pages 322--337, 05 2017.

\bibitem{moro2018discovering}
A.~Moro, B.~Garbinato, and V.~Chavez-Demoulin.
\newblock Discovering demographic data of users from the evolution of their
  spatio-temporal entropy, 2018.

\bibitem{pentland2009reality}
A.~Pentland.
\newblock Reality mining of mobile communications: Toward a new deal on data.
\newblock {\em The Global Information Technology Report 2008--2009}, 1981,
  2009.

\bibitem{thlab-sigcomm2009-20120715}
A.-K. Pietilainen and C.~Diot.
\newblock {CRAWDAD} dataset thlab/sigcomm2009 (v. 2012-07-15).
\newblock Downloaded from \url{https://crawdad.org/thlab/sigcomm2009/20120715},
  July 2012.

\bibitem{Shokri2011QuantifyingLP}
R.~Shokri, G.~Theodorakopoulos, J.-Y.~L. Boudec, and J.-P. Hubaux.
\newblock Quantifying location privacy.
\newblock {\em 2011 IEEE Symposium on Security and Privacy}, pages 247--262,
  2011.

\bibitem{Shoval2008}
N.~Shoval, G.~K. Auslander, T.~Freytag, R.~Landau, F.~Oswald, U.~Seidl, H.-W.
  Wahl, S.~Werner, and J.~Heinik.
\newblock The use of advanced tracking technologies for the analysis of
  mobility in alzheimer's disease and related cognitive diseases.
\newblock {\em BMC Geriatrics}, 8(1):7, Mar 2008.

\bibitem{unm-blebeacon-20190312}
D.~Sikeridis, I.~Papapanagiotou, and M.~Devetsikiotis.
\newblock {CRAWDAD} dataset unm/blebeacon (v. 2019-03-12).
\newblock Downloaded from \url{https://crawdad.org/unm/blebeacon/20190312},
  Mar. 2019.

\bibitem{tang2017efficient}
B.~Tang, M.~L. Yiu, K.~Mouratidis, and K.~Wang.
\newblock Efficient motif discovery in spatial trajectories using discrete
  fr{\'e}chet distance.
\newblock EDBT, 2017.

\bibitem{thomason2016identifying}
A.~Thomason, N.~Griffiths, and V.~Sanchez.
\newblock Identifying locations from geospatial trajectories.
\newblock {\em Journal of Computer and System Sciences}, 82(4):566--581, 2016.

\bibitem{thorpe2006dog}
R.~J. Thorpe, E.~M. Simonsick, J.~S. Brach, H.~Ayonayon, S.~Satterfield, T.~B.
  Harris, M.~Garcia, S.~B. Kritchevsky, A.~Health, and B.~C. Study.
\newblock Dog ownership, walking behavior, and maintained mobility in late
  life.
\newblock {\em Journal of the American Geriatrics Society}, 54(9):1419--1424,
  2006.

\bibitem{tran2012next}
L.~H. Tran, M.~Catasta, L.~K. McDowell, and K.~Aberer.
\newblock Next place prediction using mobile data.
\newblock In {\em Proceedings of the Mobile Data Challenge Workshop (MDC
  2012)}, number CONF, 2012.

\bibitem{Vhaduri2016CooperativeDO}
S.~Vhaduri and C.~Poellabauer.
\newblock Cooperative discovery of personal places from location traces.
\newblock {\em 2016 25th International Conference on Computer Communication and
  Networks (ICCCN)}, pages 1--9, 2016.

\bibitem{Wang2007SimultaneousLM}
C.-C. Wang, C.~E. Thorpe, S.~Thrun, M.~Hebert, and H.~F. Durrant-Whyte.
\newblock Simultaneous localization, mapping and moving object tracking.
\newblock {\em I. J. Robotics Res.}, 26:889--916, 2007.

\bibitem{Wang2017InferringDA}
P.~Wang, F.~Sun, D.~Wang, J.~Tao, X.~Guan, and A.~Bifet.
\newblock Inferring demographics and social networks of mobile device users on
  campus from ap-trajectories.
\newblock In {\em WWW}, 2017.

\bibitem{xu2017trajectory}
F.~Xu, Z.~Tu, Y.~Li, P.~Zhang, X.~Fu, and D.~Jin.
\newblock Trajectory recovery from ash: User privacy is not preserved in
  aggregated mobility data.
\newblock In {\em Proceedings of the 26th International Conference on World
  Wide Web}, pages 1241--1250. International World Wide Web Conferences
  Steering Committee, 2017.

\bibitem{yan2013diversity}
X.-Y. Yan, X.-P. Han, B.-H. Wang, and T.~Zhou.
\newblock Diversity of individual mobility patterns and emergence of aggregated
  scaling laws.
\newblock {\em Scientific reports}, 3:2678, 2013.

\bibitem{yang2013fine}
D.~Yang, D.~Zhang, Z.~Yu, and Z.~Yu.
\newblock Fine-grained preference-aware location search leveraging crowdsourced
  digital footprints from lbsns.
\newblock In {\em Proceedings of the 2013 ACM international joint conference on
  Pervasive and ubiquitous computing}, pages 479--488. ACM, 2013.

\bibitem{yang2015modeling}
D.~Yang, D.~Zhang, V.~W. Zheng, and Z.~Yu.
\newblock Modeling user activity preference by leveraging user spatial temporal
  characteristics in lbsns.
\newblock {\em IEEE Transactions on Systems, Man, and Cybernetics: Systems},
  45(1):129--142, 2015.

\bibitem{Zhang2018AGF}
B.~Zhang, Y.~Shen, Y.~Zhu, and J.~Yu.
\newblock A gpu-accelerated framework for processing trajectory queries.
\newblock {\em 2018 IEEE 34th International Conference on Data Engineering
  (ICDE)}, pages 1037--1048, 2018.

\bibitem{zhao2019walking}
K.~Zhao, Z.~Tu, F.~Xu, Y.~Li, P.~Zhang, D.~Pei, L.~Su, and D.~Jin.
\newblock Walking without friends: Publishing anonymized trajectory dataset
  without leaking social relationships.
\newblock {\em IEEE Transactions on Network and Service Management}, 2019.

\bibitem{geolifepaper}
Y.~Zheng, X.~Xie, and W.-Y. Ma.
\newblock Geolife: A collaborative social networking service among user,
  location and trajectory.
\newblock {\em IEEE Data Eng. Bull.}, 33:32--39, 2010.

\bibitem{zhong2015you}
Y.~Zhong, N.~J. Yuan, W.~Zhong, F.~Zhang, and X.~Xie.
\newblock You are where you go: Inferring demographic attributes from location
  check-ins.
\newblock In {\em Proceedings of the eighth ACM international conference on web
  search and data mining}, pages 295--304. ACM, 2015.

\bibitem{Zhong2015YouAW}
Y.~Zhong, N.~J. Yuan, W.~Zhong, F.~Zhang, and X.~Xie.
\newblock You are where you go: Inferring demographic attributes from location
  check-ins.
\newblock In {\em WSDM}, 2015.

\bibitem{zhou2004discovering}
C.~Zhou, D.~Frankowski, P.~Ludford, S.~Shekhar, and L.~Terveen.
\newblock Discovering personal gazetteers: an interactive clustering approach.
\newblock In {\em Proceedings of the 12th annual ACM international workshop on
  Geographic information systems}, pages 266--273. ACM, 2004.

\end{thebibliography}
}}

\end{document}